\documentclass[prd,tightenlines,nofootinbib,amsfonts,amssymb,amsmath,color,11pt]{article}
\usepackage[utf8]{inputenc}
\usepackage[affil-it]{authblk}
\usepackage{amsmath}
\usepackage{amssymb}
\usepackage{hyperref}
\usepackage{geometry}
\usepackage{color}
\usepackage{graphicx}
\usepackage{subfig}

\def\be{\begin{equation}} 
\def\ee{\end{equation}}
\def\bea{\begin{eqnarray}}
\def\eea{\end{eqnarray}}

\begin{document}

\title{\bf Primordial gravitational waves in supersolid inflation}
\author[1]{Angelo Ricciardone\thanks{angelo.ricciardone@uis.no}}
\author[2]{Gianmassimo Tasinato\thanks{g.tasinato@swansea.ac.uk}}

\affil[1]{Faculty of Science and Technology, University of Stavanger, 4036, Stavanger, Norway}
\affil[2]{Department of Physics, Swansea University,
Swansea, SA2 8PP, U.K.}

\date{\today}
 \maketitle

 \abstract{
  Supersolid inflation is a class of inflationary theories that simultaneously breaks  time and space reparameterization invariance during inflation, with  distinctive
features for the dynamics of cosmological fluctuations. We investigate 
concrete  realizations of such a scenario, including non-minimal couplings between gravity and the fields driving inflation. We
focus in particular on the dynamics of primordial  gravitational  waves
and discuss how  their properties depend on the pattern of symmetry breaking  that we consider.
Tensor modes can have a blue spectrum, and for the first time we build  models in which the squeezed limit of 
primordial tensor bispectra can be parametrically enhanced  with respect to standard single-field scenarios. At leading order in a  perturbative expansion, the tensor-to-scalar ratio depends only on the parameter controlling the breaking of space-reparameterization. 
It  is independent from the quantities controlling the breaking of time-reparameterization, and this represents a difference
with respect to standard single-field inflationary models.}

 \bigskip
\bigskip
 
\section{Introduction}
\label{sec:Intro}

The simplest approach to inflation couples gravity with a single scalar field  with a flat  potential. The scalar
field dynamics varies from model to model, but a common feature of all single-field inflationary models is the breaking of
time-reparameterization of de Sitter space during inflation. This suggests that  predictions of different  inflationary scenarios 
can be understood in terms of the language of effective field theory  applied to cosmology. Indeed, in  appropriate regimes, 
 the  curvature  perturbation ${\cal R}$  can be related with the
   Goldstone boson of the broken time-reparameterization of de Sitter
 space.
  This is 
 the spirit of the effective field theory of inflation (EFTI), started with the work~\cite{Cheung:2007st} (see
 also \cite{Weinberg:2008hq}, and  e.g.~\cite{Piazza:2013coa} for a review).

One of the most interesting lessons of the EFTI is a new perspective to inflation, that can be seen as a symmetry breaking process.  
Such viewpoint naturally motivates the exploration of more general symmetry breaking patterns, besides the minimal one which breaks time-reparameterization only. For example, one can consider models which  break also {\it space-reparameterization} during inflation. This is a possibility studied in systems 
with vector fields~\cite{Golovnev:2008cf}  or in models with scalars as solid inflation~\cite{Endlich:2012pz,Gruzinov:2004ty}. Vector models of inflation are delicate since, in the minimal  set-ups, 
 the longitudinal vector polarization  becomes a ghost around accelerating spacetimes~\cite{Himmetoglu:2008zp}. This can be cured in scenarios as 
 gauge or chromo-natural inflation~\cite{Maleknejad:2011jw,Maleknejad:2011sq,Adshead:2012kp}, or in
  models where vectors are spectator fields: see e.g. the review~\cite{Maleknejad:2012fw}. The study of cosmological
 perturbations in vector models leads to interesting features, as  possible breaking of isotropy in the scalar power spectrum, or a direction dependent squeezed limit of the
 scalar bispectrum: see e.g.~\cite{Bartolo:2012sd}. 
 Also  models with scalar fields can lead to similar features. 
  In solid inflation~\cite{Endlich:2012pz}, a set of  three scalar fields interact derivatively, and spontaneously breaks   
   space-reparameterization  during inflation. The scalar curvature perturbation is
   characterized by  a single dynamical non-adiabatic mode, which can be thought as corresponding to a phonon propagating
  in the `solid' inflationary medium. Such set-up has several distinctive observational features: among others,
    a blue spectrum for tensor
  modes,  and a direction dependent squeezed limit for the scalar  three-point function
  (with different angular dependence with respect to the aforementioned vector models).

  In this work, inspired  by  an EFTI perspective, we consider  a scenario of supersolid inflation, in which 
  a  set of four   scalar fields
   breaks both time and space reparameterizations
 during inflation (and the name supersolid is borrowed from  condensed matter nomenclature~\cite{Son:2005ak}).
  A similar scenario has already been considered in~\cite{Bartolo:2015qvr}, 
 showing that the dynamics of curvature perturbations has interesting features: the scalar three-point function has a more
 general direction-dependent squeezed limit, which interpolates between the results of vector and solid inflation, a blue spectrum for  
  tensor modes potentially detectable by future interferometers, like the Laser Interferometer Space Antenna (LISA)~\cite{AmaroSeoane:2012km} and an enhanced three-point function for graviton-scalar-scalar fluctuations.

We reconsider the scenario of supersolid inflation adding some key ingredients, as  a non-minimal coupling between
 the scalar fields and the curvature, with the specific aim to point out  new features with respect to the primordial tensor power spectrum. 
 We   build concrete models where the dynamics of cosmological
 perturbations is   straightforward to handle, and lead to interesting consequences
  for the properties  of the tensor spectrum,  which make them distinguishable from 
  single-field scenarios of inflation.   The main distinctive results are a parametrically enhanced
     tensor  non-Gaussianity  peaked in a squeezed configuration, and the identification of an interesting
     corner in parameter space where the tensor-to-scalar ratio is  only controlled by the parameters which break
     space translation during inflation (and not by the slow-roll parameter $\epsilon_H\,=\,-\dot{H}/H^2$, which controls the breaking of time-reparametrization).     
  For simplicity, we do {\it not}  study here the most general action compatible with our
  requirements, but we focus on the simplest scenarios with the properties we intend to investigate.
     
  We summarize our results in the following bullet points:
   \begin{itemize}
   \item We start with Section~\ref{sec-sys}, presenting the scalar-tensor action we build, with a new  non-minimal
     couplings between the scalars and curvature. Inflationary models of supersolid inflation 
     are conveniently 
   described in terms of three dimensionless parameters, which control respectively the breaking of 
   time-reparameterization, space-reparameterization and de Sitter invariance during inflation.  We present
   two explicit models
   of inflation, one corresponding to a system in  pure de Sitter space, the other to a model of power law inflation. 
    \item In Section~\ref{sec:tensdyn} we investigate the dynamics of tensor modes. For the first time in scenarios
    breaking space-reparameterization, we  calculate the tensor action up to third order in perturbations. 
    At second order in a perturbative expansion in the fluctuations, we find that tensors have both a mass different from zero, and a sound speed 
different than one.
 We compute the tensor power spectrum, confirming that
 tensor fluctuations can have a blue spectrum when the underlying geometry deviates from de Sitter~\cite{Endlich:2012pz}.
 At third order, the tensor action acquires contributions that are different  from 
the ones characterising single-field inflationary models. We compute the 
 corresponding contributions to tensor non-Gaussianity, finding a possible parametric
  enhancement  of tensor bispectra in their squeezed limit with respect to single-field scenarios.
 \item Section~\ref{sec:scalar-fluc} discusses the dynamics of scalar fluctuations. In a set-up as supersolid inflation, we expect
 two dynamical scalar modes: one associated with the breaking of time translations,
   the other with the breaking of space-reparameterizations. In the
 concrete models we analyse, the couplings among these two scalars can lead to an intricate coupled dynamics.
 Interestingly, we identify a corner in parameter space   where the analysis simplifies considerably: we consider
 an 
 expansion 
  at leading and next to leading   order
  in the parameter breaking time-reparameterization -- while we  keep an arbitrary 
 size for the parameter breaking space-reparameterization. At leading (zeroth) order
  in this expansion,  we find that  
 only one scalar mode propagates, related  to the comoving    
   curvature fluctuation ${\cal R}$, and the amplitude of its power spectrum
 is independent from quantities controlling the breaking of time-reparameterization. As an  interesting consequence,
  in these scenarios the tensor-to-scalar ratio only depends on the parameter breaking space-reparameterization contrarily to what usually happens in standard models of inflation driven by a single field.
 We discuss some  ramifications of this feature for the effective field theory of inflation and the Lyth bound. 
     \end{itemize}
We conclude in Section~\ref{sec:conc} where we summarize our results and discuss possible future developments.  Appendix~\ref{app-A}
contains some useful technical details.

\section{System under consideration}
\label{sec-sys}

We consider a  system of four scalar fields: $\phi$ and   $\sigma^I$, $I\,=\,1,2,3$
 whose  {\it vacuum expectation values} ({\it vevs}) spontaneously break all isometries of de Sitter space during inflation.  We wish to study the distinctive properties of fluctuations around specific configurations, focussing in particular on the  consequences  of  our symmetry breaking pattern on the spectrum and bispectrum for cosmological  fluctuations.

The explicit example of solid inflation~\cite{Endlich:2012pz,Gruzinov:2004ty} shows how to build an action for a system of three scalar fields $\sigma^I$, $I\,=\,1,2,3$, with background values depending on the space coordinates. In such a system the
scalar {\it vevs}  spontaneously break space-reparameterization, but  preserve the background isotropy and homogeneity of spacetime thanks to the following  global 
 symmetries 
 \be \label{int-sym}
\sigma^I\to O^I_J\,\sigma^J\hskip1cm,\hskip1cm \sigma^I\to \sigma^I+c^I\,,
\ee
with $O^I_J$ belonging to $SO(3)$, and $c^I$ constants. These internal global symmetries maintain the spatial rotational and translational invariance  of the background geometry during inflation. This is similar, in spirit, to the approximate shift symmetry $\phi\,\to\,\phi+const.$ usually required in single-field models of inflation for ensuring a nearly flat potential. 

Here  we use the same approach of solid inflation, but we add few key ingredients
which allow us  to explore new aspects of the consequences of these scenarios for primordial tensor modes.
 First, we include a scalar field $\phi$ with a time-dependent background profile, which couples non-derivatively to the  $\sigma^I$'s (similar scenarios were analysed in~\cite{Koh:2013msa}). The scalar $\phi$ can be thought as the standard inflaton field that breaks time-reparameterization invariance.
Second, we add a (direct) derivative coupling of the scalars $\sigma^I$ to curvature, so that the scalar system is non-minimally coupled to gravity. This is a  possibility that we
discuss for the first time in the context of theories which  spontaneously break
space-reparameterization, and that -- as we shall see  -- has interesting consequences
especially for the dynamics of tensor modes. Hence, our model is composed by four scalars that spontaneously break both time and space-reparameterization invariance during inflation.

The scalar Lagrangian density non-minimally coupled with gravity that we build is the following 
\bea
{\cal L}_{scal}\,=\,\frac12 \left( \partial \phi\right)^2+V(\phi) +q_1\, M_{Pl}^4\,
\frac{f_1(\phi)}{2}\,\delta_{IJ}\,\partial_\mu  \sigma^I\,\partial^\mu \sigma^J
+q_2\,M_{Pl}^2\,
{f_2(\phi)}\,
G^{\mu\nu} \,\partial_\mu \sigma^I\,\partial_\nu \sigma^J\,\delta_{IJ}\,,\nonumber\\
\label{scac1}\eea
where $q_1$, $q_2$ are constant dimensionless parameters, the functions $f_1(\phi)$ and $f_2(\phi)$ of the scalar $\phi$ are not specified for the moment --
 we will choose them conveniently according to the model examined --  $G^{\mu\nu}$ is the Einstein tensor, the Greek indices $\mu,\nu=\,0,\,1,\,2,\,3$
 denote spacetime coordinates, while $I,\,J=\,1,\,2,\,3$ are $\it{internal}$ indices. We use a mostly plus convention for the spacetime metric. \\
 
The previous scalar Lagrangian is the minimal one with the features we wish to investigate:
\begin{itemize}
 \item[-] It describes a minimal set of four scalar fields, able to spontaneously break time and space reparameterization symmetries, by means
 respectively of the fields $\phi$ and $\sigma^I$. 
  The scalars $\sigma^I$
 satisfy the internal symmetries~\eqref{int-sym}. Additional operators with the same properties can also be added (as done in~\cite{Koh:2013msa}) which involve 
  derivative self-couplings of the scalars $\sigma^I$. On the other hand,
 we checked   that, although  they can affect our results for the dynamics of scalar fluctuations, they  do not qualitatively change our findings for what concern the features of the tensor spectrum; hence for simplicity we do not include them in this work. 


 \item[-] For the first time, we consider a specific non-minimal coupling to curvature of models of supersolid inflation, through the operator proportional to $q_2$ in Eq.~\eqref{scac1} which includes the Einstein tensor.   We do {\it not} intend to systematically
 study all the possible non-minimal couplings with gravity in supersolid inflation. Instead we focus on the simplest
 coupling compatible with the internal symmetries \eqref{int-sym}, and with  
  distinctive consequences for the dynamics of tensor modes.  This new operator proportional to $q_2$ will play a key role in characterising the dynamics of tensor fluctuations both at level of the power spectrum and of the bispectrum. 
    Such coupling between Einstein tensor and derivatives of scalars is related with a (multifield) version of quartic Horndeski~\cite{Horndeski:1974wa}, or to the FabFour~\cite{Copeland:2012qf}. In the context of single-field inflation, the application of Horndeski scalar-tensor theory of gravity, including non-minimal couplings with curvature conceptually similar to ours, was started in~\cite{Kobayashi:2010cm,Kobayashi:2011nu}.  The structure of this action is reminiscent of an $f(\phi) F^2$ scalar-vector model \cite{Bartolo:2012sd,Watanabe:2009ct,Abolhasani:2015cve}: indeed, also in our case as in~\cite{Bartolo:2012sd}, we  choose appropriately the functions $f_1$ and $f_2$ in Eq.~\eqref{scac1}  in order to find interesting cosmological dynamics. 
\end{itemize}
To the previous scalar-tensor Lagrangian, we add the standard  Einstein-Hilbert term for gravity
 \be
S_{EH}\,=\,\frac{M_{Pl}^2}{2}\,\int\,d^4 x\,\sqrt{-g}\,R\,,
\ee
where $M_{Pl}$ is the reduced Planck mass, and we can finally express the total action as
\be \label{eq:tot-ac1}
S_{tot}\,=\,S_{EH}-\int d^4 x \sqrt{-g} \,{\cal L}_{scal}\,.
\ee
We consider a Friedman-Lemaitre-Robertson-Walker (FLRW) line element for the background metric
\be
d s^2\,=\,- d t^2+a^2(t) \,d \vec x^2\,,
\ee
where $a(t)$ is the scale factor. The positive dimensionless parameter
\be
\epsilon_H\,\equiv\,-\frac{\dot H}{H^2}\,\ge\,0\,,
\ee
where $H\,=\,\dot a/a$ is the Hubble parameter, accounts for any departure from de Sitter space (maintaining a FLRW Ansatz), $\epsilon_H=0$ corresponding
to a pure de Sitter spacetime. 

Concerning the scalar fields,
we adopt the following Ansatz for the background profiles
\bea
\phi&=&\phi(t) \label{ansp1}\,,
\\
\sigma^I&=&\lambda_0\,x^I  \label{anss1}\,,
\eea
which spontaneously break time and space reparameterization invariance respectively. Indeed, the operations of sending $t\to t+const.$ and $x^I\to x^I+const.$ do not leave invariant    the scalar background profiles. 
The dimensionless parameter $\lambda_0$ controls the breaking of space-reparameterization: notice that we adopt a simple linear Ansatz for the $\sigma^I$. We do not
instead specify the profile for $\phi$, the field that homogenously depends on time, and breaks time-reparameterization. Such background profile will depend on the model of interest.   
 
Using Ansatz~\eqref{ansp1}, \eqref{anss1}, the background equations of motion associated with action~\eqref{eq:tot-ac1} can be expressed as 
\bea \label{fri1}
H^2
\,\left( 1-\frac{q_2\,\lambda_0^2\,f_2}{a^2} \right)
&=& \frac{1}{3 M_{Pl}^2}\left[ \frac12 \dot \phi^2+V\right]+\frac{M_{Pl}^2\,q_1\,\lambda_0^2\,f_1}{2\,a^2}-
\frac{2\,q_2\,\lambda_0^2\,H\,f_2'\,\dot{\phi}}{a^2}\,,
\\\nonumber
\\
 \ddot{\phi}+3 H \dot{\phi}+V'&=&-\frac{3\,M_{Pl}^2\,\lambda_0^2}{a^2}
\left[M_{Pl}^2\,q_1\,f_1'-4\,q_2\,f_2'\,\dot{H}-6\,q_2\,f_2'\,H^2
\right]\,,
 \label{fri2}
\eea
where prime represents a derivative with respect to $\phi$, and dot a derivative with respect to physical time. All the terms proportional to $\lambda_0$ are associated with the breaking of space-reparameterization. \\
We introduce another positive dimensionless quantity
\be \label{deeph}
\epsilon_{\phi}\,\equiv\,\frac12\,\left(\frac{\dot{\phi}}{H\,M_{Pl}}\right)^2\,\ge\,0 \,,
\ee
that characterizes the amount of time-reparameterization symmetry breaking associated with a  time-dependent profile for $\phi(t)$. Hence, to sum up, the symmetry breaking parameters  at our disposal are
 
 \smallskip 
 
\begin{equation*}
\begin{array}{rl}
&{\text{\texttt{Symmetry}}} \\&{\text{\texttt{breaking}}}
 \\&{\text{\texttt{parameters}}}
\end{array}
\,=\,
\left\{
\begin{array}{rl}
\epsilon_{\phi}&\Rightarrow {\text{controls breaking of time-reparameterization symmetry}} \,,
\\
\lambda_0&\Rightarrow  {\text{controls breaking of space-reparameterization symmetry}} \,,
\\
\epsilon_{H}&\Rightarrow  {\text{controls breaking of de Sitter symmetry}} \,.
\end{array} 
%
\right.
\end{equation*}

\bigskip

 In single-clock slow-roll inflation, $\epsilon_H$ and $\epsilon_\phi$ are the same ($\epsilon_H\,=\,\epsilon_\phi$), at least in a small $\epsilon$'s limit. In our
set-up, where broken space-reparameterizations
can be included, more general possibilities can occur.  In what follows, we will be mostly interested in regimes where $\epsilon_\phi$, $\epsilon_H$ are small, while the parameter $\lambda_0$ is not necessarily a small quantity.

\smallskip

For the rest of this Section, we present two concrete
examples of background configurations which solve the previous system of equations for suitable 
choices of the parameters and the functions $f_1$ and $f_2$. These examples provide 
background solutions around which we can study cosmological perturbations in the next Sections.

\subsection{De Sitter background solution}
\label{subsec:desitt}

By choosing  appropriately the functions $f_1$ and $f_2$, we find that our system of equations admits de Sitter space as an exact solution for the metric, even if the profile for $\phi(t)$ is not constant in time, nor are the $\sigma^I$ constant in space. This implies that  we can have $\epsilon_H\,=\,0$, even if  $\epsilon_\phi$ and $\lambda_0$ are non vanishing. This shows concretely that the symmetry breaking parameters can be fairly  independent in our framework.
Concretely, we make the following choice for the functions $f_1$ and $f_2$ 
\be f_1\,=\,f_2\,=\,
\exp{\left(
\frac{\sqrt{2}\,H_0\,\phi}{M_{Pl}^2\sqrt{\kappa_0}}\right)}\,,
\ee
with $H_0$ a positive quantity  with dimension of energy,  and $\kappa_0$ a dimensionless positive constant. 
Assuming a constant scalar potential: $V\,=\,V_0$, the system admits a solution with a scale factor corresponding to
pure de Sitter space
\be
a\,=\,e^{H_0\,t}\,,
\ee
and a linear profile for the background scalar field $\phi$
\be
\phi(t)\,=\,\sqrt{2 \kappa_0}\,M_{Pl}^2\,t\,.
\ee
From the Einstein equations, evaluated in the de Sitter limit, we can see that the parameters involved are related by the conditions
\bea
\kappa_0&=&\frac{\lambda_0^2}{2}
\,\left(6\,q_2\,\frac{H_0^2}{M_{Pl}^2}-q_1\right)\,,
\eea
and the Hubble parameter $H_0$ is given by
\bea
H_0^2
&=&\frac{1}{3\,M_{Pl}^2}\, \frac{V_0+M_{Pl}^4\,q_1 \,\lambda_0^2}{1+2\,q_2\,\lambda_0^2}
\\
&=&\frac{V_0}{3\,M_{Pl}^2} -\frac{2  \,\kappa_0\,M_{Pl}^2}{3}
\,.
\eea
Since $H_0$ is constant  $\epsilon_H=0$, and the parameter
$\epsilon_\phi$ (as defined in Eq.~\eqref{deeph}) results
\be
\epsilon_{\phi}\,=\, \kappa_0\,\frac{M_{Pl}^2}{H_0^2}\,,
\ee
so it is controlled by the quantity $\kappa_0$. 
Substituting these results in the expressions for $f_1$, $f_2$, we find that these two functions are proportional to the square of the 
scale factor 
\be
f_1\,=\,f_2\,=\,a^2\,=\,e^{2\,H_0\,t}\,.
\ee
So to summarize, this configuration spontaneously break space and time reparameterization symmetries of the system 
through the fields {\it vevs}, although the background metric is de Sitter space with $\epsilon_H=0$. 

\subsection{Power law background solution}
\label{subsec:powerlaw}
Our system admits also a solution corresponding to power law inflation. 
We make the following choice for the functions and parameters involved
\bea
V&=&V_0 \,e^{-\beta \phi/M_{Pl}}\,,
\\
f_1&=&e^{(p-1) \beta \phi/M_{Pl}}\,,
\\
f_2&=&e^{p \beta \phi/M_{Pl}}\,,
\eea
with
 \bea
 V_0&=&M_{Pl}^4\left[ p\left( 3 p-1\right) + 
\lambda_0^2\left[ \left( 6 p-1\right) p\,q_2-q_1 \right]
\right]\,,
\\
\beta&=&\frac{2}{\sqrt{2 p+\left(2 \,p\, q_2 +6 \,p^2\, q_2 -q_1\right)\, \lambda_0^2 }} \,,
\label{solbsi}
\eea
and $p$ is a dimensionless constant. In order to ensure a real value of $\beta$ we impose the following inequality
\be \label{ine1}
 2p\left(1+q_2 \lambda_0^2\right)+6\,p^2\, q_2\,\lambda_0^2\,\ge\,q_1\,\lambda_0^2\,.
\ee
The solution for the  background equations reads
\bea
a&=& \,\left( M_{Pl}\,t\right)^p\,,
\\
H&=&\frac{p}{t}\,,
\\
\phi&=&\frac{2 \,M_{Pl}}{\beta}\,\ln{\left( M_{Pl}\,t\right)}\,.
\eea
The structure of these equations  is the familiar one characterising power law inflationary models (see for example~\cite{Lucchin:1984yf,Liddle:1988tb}). For what concern the symmetry breaking parameters in the power law set-up, we find that
 \bea
 \epsilon_H&\equiv&-\frac{\dot H}{H^2}\,=\,\frac{1}{p}\,,
\\
\epsilon_\phi&=& \frac{2}{\beta^2\,p^2}\,=\,\frac{2 \,p+\left(2 \,p\, q_2 +6 \,p^2\, q_2 -q_1\right)\, \lambda_0^2}{2 p^2}\,.
 \eea
 Notice that while $\epsilon_H$ is only controlled by the parameter $p$, the quantity $\epsilon_\phi$ depends also on other
 quantities. Hence in this case we can consider the two quantities $\epsilon_H$, $\epsilon_\phi$ as two positive independent parameters.

\section{Dynamics of tensor fluctuations}
\label{sec:tensdyn}
Our first  aim is to investigate the dynamics of gravitational waves around the background configurations examined in the previous Section. Tensor fluctuations represent an ubiquitous prediction of all the inflationary models and they reflect the features of the theory of gravity responsible for the accelerated expansion. Their predictions are more model-independent than scalar fluctuations and they represent a unique opportunity to provide information on the energy scale of inflation and on specific symmetry patterns that characterize the inflationary epoch. Other studies of tensor fluctuations in set-ups where space-reparameterization is broken include~\cite{Cannone:2014uqa,Cannone:2015rra}. 
  
Focusing only on tensor fluctuations (and neglecting vector and scalar perturbations) we express
the perturbed FLRW metric as 
\be \label{pmet1}
d s^2\,=\,g_{\mu \nu}\,d x^\mu\,d x^\nu\,=\,-d t^2+a^2(t)\left( \delta_{ij}+\gamma_{ij}\right)\,d x^i \,d x^j\,.
\ee
We expand the spatial metric fluctuation up to third order in the tensor mode following~\cite{Maldacena:2002vr}
\be \label{exp-gam}
\gamma_{ij}\,=\,h_{ij}+\frac12 h_{i k}\,h_{k j}+\frac16 h_{i k} h_{k l} h_{lj}\,,
\ee
where $h_{ij}$ represents a first order, transverse ($\partial_{i} h^{i}_{\,j}=0$) and traceless ($h^{i}_{\,i}=0$) tensor fluctuation.  
Indexes are contracted with 3D Kronecker  function. 
The choice \eqref{exp-gam} is particularly convenient for the expansion since it gives $\sqrt{-g}\,=\,a^3$.

The action for the tensor modes is obtained  plugging the metric~\eqref{pmet1} in the
original action~\eqref{eq:tot-ac1}, and perturbing up to desired order around the background configuration of interest. 
The structure of the action to examine is the following
\bea \label{tac1}
\, S\,=\,
\int \,d^4 x\,\sqrt{-g}\,\left[ \frac{M_{Pl}^2}{2}R
+\frac{\dot \phi^2}{2}
-V(\phi) -\lambda_0^2\,q_1\, M_{Pl}^4\,
\frac{f_1(\phi)}{2}\,\delta_{ij}\,g^{ij}
-\lambda_0^2\,q_2\,M_{Pl}^2\,
{f_2(\phi)}\,
\delta_{ij}\,G^{ij}\right]\,,
\eea
where $\phi$ is an homogeneous field only depending on time. See Appendix \ref{app-A} for an expansion of each term of this action up to third order
in tensor fluctuations.  

\subsection{Quadratic tensor action: tensor mass and sound speed}
\label{subsec:sectens}
The first distinctive signatures of our scenario appear already at quadratic order: we now show that tensors have non-vanishing mass, and a sound
speed different from unity. To study these properties we expand the action at second order in tensor fluctuations and get
\be
S_2^{(T)}\,=\,\frac{M_{Pl}^2}{8} \,\int \,d t d^3 x\,a^3\,\left( 1-\frac{q_2\,\lambda_0^2\,f_2 }{a^2}\right)\left[
\dot{h}^2_{ij}-\frac{c_T^2}{a^2}\,\left(\nabla h_{ij}\right)^2-m_h^2\, h_{ij}^2
\right]\,,
 \label{gesqa}
\ee
where an overdot denotes a derivative wrt time. The tensor speed of sound and mass for 
general choices of $f_1$, $f_2$ are
\bea
c_T^2&=&\frac{1-3  q_2\,\lambda_0^2\,f_2/a^2}{1-  q_2\,\lambda_0^2\,f_2/a^2}\,,
\\\nonumber
\\
m_h^2 &=&-
\frac{4\,\lambda_0^2}{\left(1-  q_2\,\lambda_0^2\,f_2/a^2\right)^2}
\Big[\frac{q_2}{a^2}\,
\left( 2- 3  q_2\,\lambda_0^2\,f_2/a^2\right)H^2
-\frac{M_{Pl}^2}{2}\,q_1\,\frac{f_1}{a^2}\nonumber\\
&& -\frac{q_2}{2 a^2}\,\left( 1- 9  q_2\,\lambda_0^2\,f_2/a^2\right)\,f_2'\,\dot{\phi}\,H
-\frac{q_2\,\dot \phi^2}{2\,M_{Pl}^2\, a^2} \left(f_2 -M_{Pl}^2 \,f_2''\, \left( 1-3    q_2\,\lambda_0^2\,f_2/a^2  \right) \right)
\nonumber\\ &&+\frac{q_2}{2\,a^2}\left(1-3    q_2\,\lambda_0^2\,f_2/a^2  \right)\,f_2' \,\ddot{\phi}
\Big]\,.
\eea
Due to the breaking of space-reparameterization through $\lambda_0\neq0$,
the tensor mass and sound speed can acquire values different from the standard case of single-field inflation
(i.e.  $m_h=0$, $c_T=1$).  The graviton mass is a distinctive feature associated with    
 breaking of space-reparameterization while a tensor sound speed different from one is due to the non-minimal coupling of the scalars with gravity, and to the scalar-tensor kinetic mixing.  

We can now apply the previous formulae to the two scenarios discussed in the previous Section, pure de Sitter (Section~\ref{subsec:desitt}) and power law inflation (Section~\ref{subsec:powerlaw}).  In both cases, the 
tensor sound speed is constant and it is given by
\bea \label{def-ct}
c_T^2
&=&\frac{1-3  q_2\,\lambda_0^2}{1-  q_2\,\lambda_0^2}\,,
\eea
while the Planck mass gets ``renormalised'' to a value
\be
 M_{Pl}\,\to\,\bar M_{Pl}\,\equiv\,M_{Pl}\,\sqrt{1-{q_2\,\lambda_0^2}}\,,
\label{eq:renplanck}
\ee
due to the non-minimal coupling of our scalars with gravity (recall that $q_2$, $\lambda_0$ are constant). In order to have a well-defined Planck mass, and a well-defined sound speed smaller than one, from now on we impose the following condition
\be \label{condqq2}
0\,\le\, q_2 \lambda_0^2\,\le\,1/3\,.
\ee
\smallskip
Differences among de Sitter and power law inflation arise when computing the value of the graviton mass during inflation. For the case of pure de Sitter space, the graviton mass during inflation is given by
\bea 
\frac{m_{h,\,\,deSit}^{2}}{H^2}&=&
-\frac{4\,\kappa_0}{M_{Pl}^2 \,H_0^2\,\left(1-q_2 \lambda_0^2\right) }\,,
\\
&=&-\frac{4\epsilon_\phi}{ \left(1-q_2 \lambda_0^2\right) }\le0\,,
 \label{mdsc}
\eea
so we find a negative mass squared. 
This is consistent with the Higuchi bound, which states that in pure de Sitter space the graviton mass can not
lie on the interval $0<m_h^2\le 2 H^2$~\cite{Higuchi:1986py}. We find that the graviton mass is proportional to the parameter $\epsilon_\phi$
controlling the breaking of time-reparameterization. 

On the other hand for power law inflation we find
\bea
\frac{
m_{h,\,plw}^2}{H^2}&=&-\frac{2\,\lambda_0^2}{p^2}\,\frac{2 \,(3p-2) \,p\,q_2 -q_1}{1-\lambda_0^2 q_2}\,,
\\
&=&-\frac{4\epsilon_\phi}{1-q_2 \lambda_0^2}+ {4\,\epsilon_H}\,\frac{1+3 q_2 \lambda_0^2}{1-q_2 \lambda_0^2}\,.
 \label{mpli}
\eea
Since $\epsilon_\phi$ and $\epsilon_H$ are positive, independent quantities, 
either sign of $m_{h}^2$ can be obtained. This is consistent again with Higuchi bound, since we are not in pure de Sitter space since $\epsilon_H\neq 0$. 
We have to satisfy inequality \eqref{ine1} though, which leads to an upper bound for the graviton mass during 
power law inflation
\be
\frac{
m_{h,\,plw}^2}{H^2}\,\le\,{4\,\epsilon_H}\,\frac{1+3 q_2 \lambda_0^2}{1-q_2 \lambda_0^2}\,,
\ee
 so -- if positive -- it is suppressed by a parameter $\epsilon_H$.  If we insist on a positive $m_{h}^2$,  this parameter
   has to be small in a quasi de Sitter limit (but notice that, on the other hand,  it can be large and negative, since there is no lower bound).
   Notice that the result in Eq.~\eqref{mpli} generalizes  Eq.~\eqref{mdsc}: in the limit $\epsilon_H\to0$, the former reduces to the latter. 

A non vanishing tensor mass $m_h\neq 0$ indicates that  tensor modes are {\it not} adiabatic fluctuations during
inflation. This is usually associated with scenarios where inflation is not an efficient isotropic attractor: namely, anisotropic features do not necessarily decay exponentially. This is not necessarily a bad feature, and might lead to interesting observational consequences~\cite{Bartolo:2013msa,Bartolo:2014xfa,Akhshik:2014bla,Akhshik:2014gja}.   

\bigskip

In order to compute power spectrum and bispectrum for tensor modes, 
 we need to quantize tensor fluctuations, see for example~\cite{Lyth:2009zz}. 
 Here we follow the notation of~\cite{Gao:2011vs} and we write  tensor fluctuations in Fourier space as
\be
h_{ij}(t,\vec x)\,=\,\int\frac{d^3 k}{(2 \pi)^3}
\,\tilde{h}_{ij}(t,\vec k)\,e^{i\,\vec k \cdot \vec x}\,.
\ee
The Fourier mode $\tilde h_{ij}$ can be quantized and decomposed in terms of polarization tensors, and creation/annihilation operators
 \be
 \tilde h_{ij}\,=\,\sum_s\,\left[\chi_k\,{\bf e}_{ij}^{(s)}(\vec k)\,a_{s}(\vec k)+ \chi_{-k}^*\,{\bf e}_{ij}^{*\,(s)}(-\vec k)\,a^\dagger_{s}(-\vec k)\right]\,,
 \ee
 with ${\bf e}_{ij}^{(s)}$ indicating the polarization tensor with helicity $s=\pm2$, satisfying the transverse-traceless condition $\,k_i \,{\bf e}_{ij}^{(s)}\,=\,{\bf e}_{ii}^{(s)}=\,0$.  
 We adopt the normalization conditions: ${\bf e}_{ij}^{(s)}\,{\bf e}_{ij}^{(s')}\,=\,\delta_{s s'}$. We also use the following property ${\bf e}_{ij}^{*\,(s)}(\vec k)\,=\,{\bf e}_{ij}^{(-s)}(\vec k)\,=\,{\bf e}_{ij}^{(s)}(-\vec k)$. The creation/annihilation operators satisfy the usual commutation relations $[a_{s}(\vec k),a^\dagger_{s'}(\vec k')]=(2\pi)^3\delta_{ss'}\delta^{(3)}(\vec{k}-\vec{k}')$.  
 Requiring to match the Bunch-Davies vacuum  at early time, we find that the mode function $\chi_k$ is  equal to
\be
 \chi_k\,=\,\frac{\sqrt \pi}{a}\,\sqrt{\frac{-y}{c_T}}\,H^{(1)}_{\nu}\left(-k y\right)\,,
 \ee
 where we have introduced a new time coordinate $d y\,=\,c_T\,d t/a$ and $H^{(1)}_{\nu}$ is the Hankel function of the first kind with
 \be \label{defnu}
 \nu\,=\, \frac{1}{1-\epsilon_H} \sqrt{\frac{\left(3-\epsilon_H\right)^2}{4}-\frac{\,m_h^2}{H^2}}\,.
%
 \ee
In order to be as general as possible we focus on the power law inflationary solution described in Section~\ref{subsec:powerlaw}.
 The model around   de Sitter space is a special case of the one discussed here,  with $\epsilon_H=0$. In the power law set-up we have
 \be
 y\,=\,-\frac{c_T}{1-\epsilon_H}\,\frac{1}{a\,H}\,,
 \ee
and now we are ready to compute the primordial power spectrum. The tensor two-point function is defined as
 \be
 \langle \tilde  h_{ij} (\vec k)  \,\tilde h_{lm}(\vec k') \rangle \,=\,\left(2 \pi \right)^3\,\delta^{(3)} (\vec k+
 \vec k')\,|\chi_k |^2\,\Pi_{ij,\,lm}(\vec k)\,,
 \ee
 where we have introduced 
 \be
 \Pi_{ij,\,lm}(\vec k)\,=\,\sum_s\,{\bf e}_{ij}^{(s)}(\vec{k})\,{\bf e}_{lm}^{*(s)}(\vec{k})\,.
 \ee
Starting from the quantity
  \be
  {\cal P}_{ij,\,lm}\,=\,|\chi_k |^2\,\Pi_{ij,\,lm}(\vec k)\,,
  \ee
 we can define the power spectrum for tensor fluctuations as ${\cal P}_{h}(k)\,=\,\left(k^3/2 \pi^2\right)\,{\cal P}_{ij,\,ij}$ and, focussing on large scales, 
 we obtain
 \be
 {\cal P}_h(k)\,=\,\frac{H^2}{\bar{M}_{Pl}^2}\,
 \frac{
 \left(1-\epsilon_H\right)^2\,2^{2 \nu-2}}{ \pi^2\,c^3_T} \,\left(\frac{k}{k_*}\right)^{3-2\nu}\,,
 \label{eq:tensps}
 \ee
 where $c_T\, k_*\,=\,a\, H$ is the scale at the horizon exit, $c_T$ is the tensor sound speed and $\bar{M}_{Pl}$ is the renormalized Planck mass~\eqref{eq:renplanck}. Considering a constant tensor speed of sound we can easily compute the tensor spectral index
 \bea \label{ntdef1}
 n_T&\equiv&\frac{d\,\ln{ {\cal P}_h}}{d\,\ln{k}}\,\Big|_{ k\,=\,k_*}\,=\,-2\,\epsilon_H+{3-2\nu}\,,
 \\
 &\simeq&-2\,\epsilon_H+\frac{2}{3}\,\frac{m_h^2}{H^2}\,,
 \\
 &=&\frac{1}{1-q_2 \lambda_0^2}\left[-4 \epsilon_{\phi}+2\,\epsilon_H\,
 \left(1+7q_2 \lambda_0^2\right)\right]\,,
 \eea
 where in the last two lines of the previous equations we focussed on the limit of small symmetry breaking parameters $\epsilon_\phi$, $\epsilon_H$, which 
 also imply small graviton mass.   In our set-up of power law inflation, where $\epsilon_H$, $\epsilon_\phi$, $\lambda_0$ are independent parameters, we can obtain $n_T\ge 0$  (but
 small if the $\epsilon$'s are small), hence a blue spectrum, that is a distinctive signature of  scenarios like
 solid inflation \cite{Endlich:2012pz}. 
In standard single-field slow-roll inflation, instead, one has $\lambda_0=0$, $\epsilon_\phi=\epsilon_H$, hence $n_T\,=\,-2
 \epsilon_H$ (red spectrum).

\smallskip

Besides models which break space-reparameterization and lead to an effective graviton mass during inflation, other systems which produce a blue
spectrum for primordial tensor modes include scenarios 
 with particle production during inflation~\cite{Green:2009ds,Anber:2009ua,Barnaby:2010vf}, and set ups where tensors are sourced by spectator fields~\cite{Biagetti:2013kwa,Biagetti:2014asa}.  
 See~\cite{Guzzetti:2016mkm} for a comprehensive review, and~\cite{Bartolo:2016ami}  for a related
  analysis of the perspectives of future detection of primordial tensor modes 
with interferometers, like LISA. An advantage of our framework is that we only use the same fields that drive
inflation, hence, by construction, we avoid the delicate backreaction issues which must be taken into account by other scenarios which make
use of additional fields besides the inflaton.

\smallskip

 It is important to emphasize that we do have both non-standard tensor sound speed and
 graviton mass during inflation. While in standard EFTI a combination of disformal and conformal transformations
 could bring the sound speed to one, as explained in~\cite{Creminelli:2014wna} (but see also~\cite{Fumagalli:2016afy}), in our case those operations would have implications
 for the quadratic term in the tensor action proportional to the graviton mass, and {would change the scale dependence of the tensor spectrum.}

Let us show this explicitly: we assume -- as happens for our concrete example -- that the tensor  sound speed is constant. 
The combination of disformal plus conformal transformations which allows one to set the tensor sound speed to one is~\cite{Bekenstein:1992pj}
 \be
  g_{\mu\nu}\,\to\,c_T^{-1} \left[ g_{\mu \nu}+\left(1-c_T^2\right)\,n_\mu n_\nu\right]\,,
 \ee
 with $n_\mu \propto \partial_\mu \phi$. This transforms the spacetime metric to $d s^2\,=\,-c_T\,d t^2+c_T^{-1}\,a^2\,d \vec{x}^2$. We further rescale
  the time coordinate and the scale factor as $t\,\to\,c_T^{-1/2} t$ and $a\,\to\,c_T^{1/2}\,a$ to express  the metric in a standard FLRW form. The tensor
  action gets transformed to
  \bea
  S_T&=&\int dt d^3 x\, a^3\left[\dot{h}_{ij}^2-\frac{c_T^2}{a^2}\left( \nabla h_{ij}\right)^2-m_h^2 {h}_{ij}^2\right]
 \\
 &\to&  c_T^2\,\ \int d t d^3 x \, a^3\,\left[\dot{h}_{ij}^2-\frac{1}{a^2}\left( \nabla h_{ij}\right)^2-\frac{m_h^2}{c_T} {h}_{ij}^2\right]\,,
  \eea
  hence, after the transformation, the tensor sound speed is one, but the graviton mass gets enhanced (if $c_T<1$). In
   such a frame, a large graviton mass would lead to a  {large value of the tensor tilt   -- see Eqs.
    \eqref{defnu} and \eqref{ntdef1}.} 	
\subsection{Cubic tensor action and tensor non-Gaussianity}
\label{subsec:cubictensor}
At the same level of scalar, tensor non-Gaussianity represents a powerful tool to discriminate among inflationary models and in particular the study of the bispectrum shape and amplitude can open the possibility to test consistency relations that are a valid tool to test symmetries in the early universe.\\
In this Section we analyze tensor non-Gaussianity, a subject that so far has not been studied much in the literature. Other works discussing this topic include e.g.~\cite{Maldacena:2011nz,Soda:2011am,Cook:2013xea,Akita:2015mho}.
 We study the subject for the first time in scenarios breaking space-reparameterization revealing
 additional and distinctive features of this class of models.
  
  In order to compute the three-point function we expand the action~\eqref{eq:tot-ac1} at third order 
in tensor fluctuations using~\eqref{exp-gam}.
For both the set-ups of pure de Sitter and power law inflation, we find the same structure for the 
tensor action at third order 
 \bea
 S_3^{(T)}&=&
\frac{\bar{M}_{Pl}^2}{4}\,
\int\,d t \,d^3 x\,
a^3\,\left[
{\cal T}_1+{\cal T}_2 
\right]\nonumber
\\ 
&=&
\frac{\bar{M}_{Pl}^2}{4}\,
\int\,d t \,d^3 x\,
a^3\,\left[ 
\frac{1}{ a^2}\,\frac{\left(1-5 q_2 \lambda_0^2\right)}{\left(1-q_2 \lambda_0^2\right)\,}\,h_{ij} h_{n m}\left( \partial_j \partial_n \,h_{im}-\frac12 \partial_i \partial_j\,h_{mn}\right)
+\frac{m_h^2}{6}\,h_{ij} h_{jm} h_{mi}\right]\,,
\nonumber\\
\label{eq:cubicaction}
\eea
with the renormalised Planck mass given in Eq.~\eqref{eq:renplanck}. The third order action is composed by two terms:

\smallskip

\noindent
$\bullet$ 
{\bf The first contribution ${\cal T}_1$} depends on spatial derivatives of the tensor $h_{ij}$, and 
 is identical in structure to the one found
in General Relativity (GR) around de Sitter space. The only difference is the constant overall factor.
We compute  the three-point function for  the  tensor mode $h_{ij}$ associated with the ${\cal T}_1$ piece using the in-in formalism~\cite{Maldacena:2002vr, Weinberg:2005vy}
  \begin{equation}
  \langle \tilde{h}_{i_{1}j_{1}}(\vec{k}_1)\tilde{h}_{i_{2}j_{2}}(\vec{k}_2)\tilde{h}_{i_{3}j_{3}}(\vec{k}_3)\rangle =-i \int_{-\infty}^0 d\,t'
   \left \langle \left[\tilde{h}_{i_{1}j_{1}}(t,\vec{k}_1)\tilde{h}_{i_{2}j_{2}}(t,\vec{k}_2)\tilde{h}_{i_{3}j_{3}}(t,\vec{k}_3), H_{\rm{int}}(t')\right]\right \rangle \;,
  \end{equation}
 where $H_{\rm{int}}$ is the interaction Hamiltonian obtained by the cubic action~\eqref{eq:cubicaction}. We obtain
\be
\langle \tilde h_{i_1 j_1}(\vec k_1)  \tilde h_{i_2 j_2}(\vec k_2)  \tilde h_{i_3 j_3}(\vec k_3)\rangle_{{\cal T}_1}\,=\,\frac{\left(1-5 q_2 \lambda_0^2 \right)}{\left(1-3 \lambda_0^2 q_2\right)} \,\times \,(2 \pi)^7\,\delta^{(3)}\left(\vec k_1+\vec k_2 +\vec k_3 \right)\,{\cal P}_h^2\,\frac{
{\cal A}^{{\cal T}_1}_{i_1 j_1\,i_2 j_2\,i_3 j_3}}{k_1^3\,k_2^3\,k_3^3}\,,
\ee
where, following~\cite{Gao:2011vs}, we have introduced the non-Gaussian amplitude ${\cal A}^{{\cal T}_1}_{i_1 j_1\,i_2 j_2\,i_3 j_3}$, that results  equal to
\begin{eqnarray}
{\cal A}_{i_1j_1i_2j_2i_3j_3}^{{\cal T}_1} &=&
\tilde{\cal A}
\Big\{
\Pi_{i_1j_1,ik}(\vec{k}_1)\Pi_{i_2j_2,jl}(\vec{k}_2)
\left[
k_{3k}k_{3l}\Pi_{i_3j_3,ij}(\vec{k}_3)
-\frac{1}{2}k_{3i}k_{3k}\Pi_{i_3j_3,jl}(\vec{k}_3)
\right]
\nonumber\\
&&
\hskip0.5cm +5 ~{\rm perms.}~{\rm of}~ 1, 2, 3
\Big\}\,,
\end{eqnarray}
with $\tilde{\cal A}$ given by
\begin{eqnarray}
\tilde {\cal A}(k_1,k_2,k_3)=
-\frac{K}{16}\biggl[
1-\frac{1}{K^3}\sum_{i\neq j}k_i^2k_j-4\frac{k_1k_2k_3}{K^3}
\biggr]\,,\;\;\;
\end{eqnarray}
and $K=k_1+k_2+k_3$.
 
\begin{figure}[t!]
  \begin{center}
    \includegraphics[keepaspectratio=true,height=45mm]{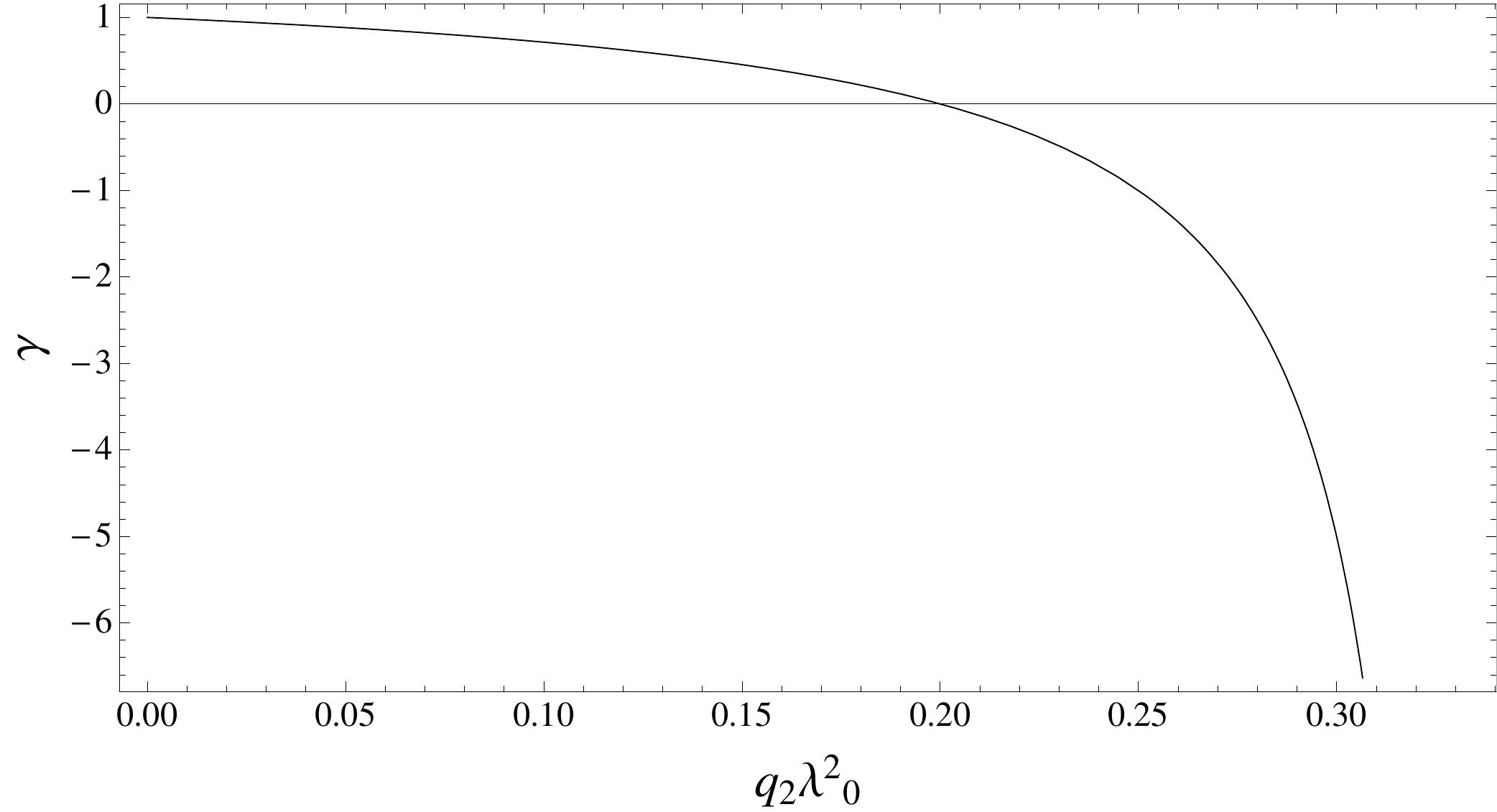}
  \end{center}
  \caption{Evolution of $\gamma$, defined in Eq.~\eqref{gadef1} in the range of values allowed by Eq.~\eqref{condqq2}. As explained in the text, the parameter $\gamma$ controls 
   the amplitude of the   squeezed limit of tensor bispectrum in our model, and how much it differs from the same quantity calculated in standard inflationary scenarios. 
  }%
  \label{fig:gammaq2}
\end{figure}
As anticipated, the momentum dependence for this three-point function is  equal to the GR contribution around de Sitter space found in~\cite{Gao:2011vs}, but the overall factor is different. In fact,   the amplitude of the bispectrum is proportional to the factor $\gamma$ given by   
   \be \label{gadef1}
 \gamma\,=\,\frac{\left(1-5 q_2 \lambda_0^2 \right)}{\left(1-3 q_2 \lambda_0^2\right)} \,.
 \ee
Interestingly, this factor can be well different from one. We can plot $\gamma$ versus $q_2\,\lambda_0^2 $, with the latter 
quantity varying within the range allowed by Eq.~\eqref{condqq2}: we find that when
taking the limit
 \be q_2\,\lambda_0^2  \,\to\,\frac13\,, \ee
 the 
factor $\gamma$ is large and negative,  as shown in  Fig.~\ref{fig:gammaq2}.  Notice that the amplitude of the bispectrum gets enhanced in the region where
the tensor sound speed becomes small, see Eq.~\eqref{def-ct}. 
   
\begin{figure}[t!]
  \begin{center}
    \includegraphics[keepaspectratio=true,height=60mm]{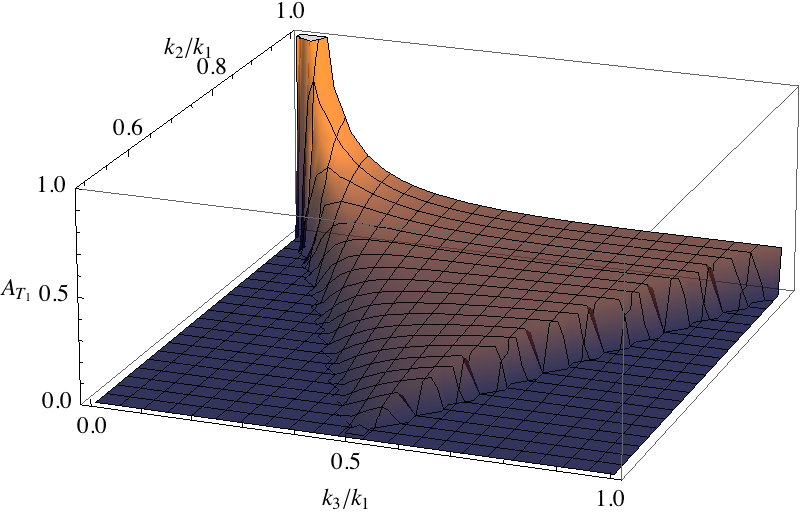}
  \end{center}
  \caption{${\cal A}_{{\cal T}_{1}}^{+++}(1, k_2/k_1, k_3/k_1)(k_2/k_1)^{2}(k_3/k_1)^{2}$
  as a function of $k_2/k_1$ and $k_3/k_1$. The plot is normalized to unity for
  equilateral configurations $k_2/k_1=k_3/k_1=1$. The contribution to the tensor bispectrum associated with operator ${\cal T}_1$ peaks in the
  squeezed limit.  }%
  \label{fig:grshape}
\end{figure}

   The tensor bispectrum we evaluated has its maximum
   contribution in the squeezed limit, as in the case of standard single-field inflationary models.   In order to appreciate this feature
   more clearly,  we focus on the
two polarization modes defined as
\begin{eqnarray}
\xi^{(s)}(\vec{k})
= \tilde h_{ij}(\vec{k}) e^{*(s)}_{ij}(\vec{k})\,.
\end{eqnarray}
We  consider the amplitude ${\cal A}^{s_1s_2s_3}$ and the shape
of the bispectrum
$\langle \xi^{s_1}(\vec{k}_1)\xi^{s_2}(\vec{k}_2)\xi^{s_3}(\vec{k}_3)\rangle$. We
have
$
{\cal A}_{{\cal T}_{1}}^{s_1s_2s_3}=
e_{i_1j_1}^{*(s_1)}(\vec{k}_1)
e_{i_2j_2}^{*(s_2)}(\vec{k}_2)
e_{i_3j_3}^{*(s_3)}(\vec{k}_3)
{\cal A}_{i_1j_1i_2j_2i_3j_3}^{{\cal T}_{1}}
$ and we obtain
\begin{eqnarray}\label{at1}
{\cal A}_{{\cal T}_{1}}^{s_1s_2s_3}&=&\gamma \tilde{\cal A}(k_1,k_2,k_3)
F_{{\cal T}_{1}}^{s_1s_2s_3}(k_1,k_2,k_3),
\end{eqnarray}
where $F_{{\cal T}_{1}}$ coincides with the standard single-field inflation results found in~\cite{Gao:2011vs}
\begin{equation}
F_{{\cal T}_{1}}^{+++}(k_1,k_2,k_3)=\frac{1}{2}\frac{K^5}{64k_1^2 k_2^2k_3^2} \bigg[K^3-4
\sum_{i\neq j}k_i^2k_j-4 k_1k_2k_3 \bigg]\,.
\end{equation}
We  plot in Fig.~\ref{fig:grshape} the amplitude of this first contribution confirming that it peaks in the squeezed limit $k_{3}\rightarrow 0$.

While the shape is the same as in standard single-field scenarios,   the amplitude is modified and can be enhanced through the factor $\gamma$, hence it might be easier to observationally detect.

\begin{figure}[t!]
  \begin{center}
    \includegraphics[keepaspectratio=true,height=60mm]{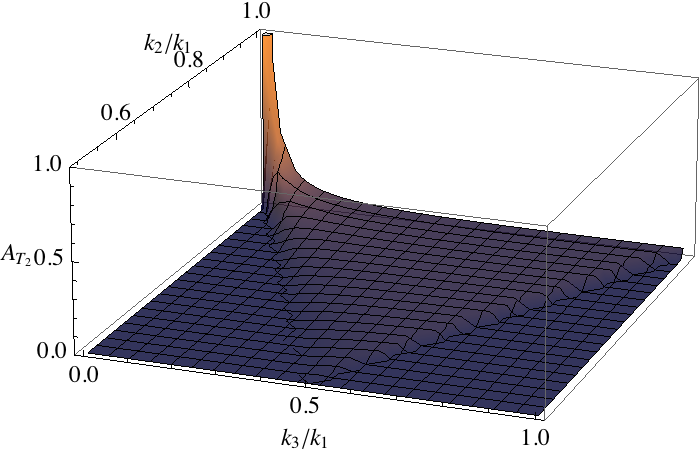}
  \end{center}
  \caption{${\cal A}_{{\cal T}_{2}}^{+++}(1, k_2/k_1, k_3/k_1)(k_2/k_1)^{2}(k_3/k_1)^{2}$
  as a function of $k_2/k_1$ and $k_3/k_1$. The plot is normalized to unity for
  equilateral configurations $k_2/k_1=k_3/k_1=1$.}%
  \label{fig:massashape}
\end{figure}

\bigskip
$\bullet$ 
{\bf The second contribution ${\cal T}_2$} is proportional to the mass of the graviton $m_h^2$ and is 
 distinctive of the scenario that we have considered. Such a contribution is expected when space-reparameterizations are broken, since there is no symmetry that prevents this term. It can be relevant in scenarios in which the size of the graviton mass $|m_h^2|$ is large (although we will not consider these cases in what follows).  Following a procedure similar to the first contribution ${\cal T}_1$, we find that 
\begin{equation}
\langle \tilde h_{i_1 j_1}(\vec k_1)  \tilde h_{i_2 j_2}(\vec k_2)  \tilde h_{i_3 j_3}(\vec k_3)\rangle_{{\cal T}_2}\,=\,\,(2 \pi)^7\,\delta^{(3)}\left(\vec k_1+\vec k_2 +\vec k_3 \right)\,{\cal P}_h^2\,\frac{
{\cal A}^{{\cal T}_2}_{i_1 j_1\,i_2 j_2\,i_3 j_3}}{k_1^3\,k_2^3\,k_3^3}\,,
\label{eq:t2}
\end{equation}
where ${\cal A}^{{\cal T}_2}_{i_1 j_1\,i_2 j_2\,i_3 j_3}$ is now given by
\begin{equation}\label{at2}
{\cal A}_{i_1j_1i_2j_2i_3j_3}^{{\cal T}_2}=
\frac{m_{h}^2}{ H^2} 
\tilde{\mathcal{A}}_{\mathcal{T}_2}\Pi_{i_1j_1, lm}(\vec{k}_1)
\Pi_{i_2j_2, mn}(\vec{k}_2)
\Pi_{i_3j_3, nl}(\vec{k}_3)\,,
\end{equation}
with
\begin{equation}
\tilde{\mathcal{A}}_{\mathcal{T}_2}(k_1,k_2,k_3)=
\frac{1}{48}\biggl[k_1k_2k_3
+\sum_{i\neq j}k_i^2k_j+(1-\gamma_{E})\sum_{i}k_i^3\biggr]\,,\;\;\;
\end{equation}
and $K=k_1+k_2+k_3$.\\
Like for the contribution $\mathcal{T}_{1}$ we have $
{\cal A}_{{\cal T}_{2}}^{s_1s_2s_3}=
e_{i_1j_1}^{*(s_1)}(\vec{k}_1)
e_{i_2j_2}^{*(s_2)}(\vec{k}_2)
e_{i_3j_3}^{*(s_3)}(\vec{k}_3)
{\cal A}_{i_1j_1i_2j_2i_3j_3}^{{\cal T}_{2}}
$ and we obtain
\begin{eqnarray}
{\cal A}_{{\cal T}_{2}}^{s_1s_2s_3}&=&\tilde{\mathcal{ A}}_{\mathcal{T}_{2}}(k_1,k_2,k_3)
F_{{\cal T}_{2}}^{s_1s_2s_3}(k_1,k_2,k_3),
\end{eqnarray}
where now $F_{{\cal{T}}_2}^{s_1s_2s_3}(k_1,k_2,k_3)$, using the properties of the polarization tensors, results
\begin{equation}
F_{{\cal T}_{2}}^{+++}(k_1,k_2,k_3)=\frac{K^3}{64k_1^2 k_2^2k_3^2} \bigg[K^3-4
\sum_{i\neq j}k_i^2k_j-4 k_1k_2k_3 \bigg]\,.
\end{equation}
In Fig.~\ref{fig:massashape} we plot the non-Gaussian amplitude correspondent to the contribution proportional to the mass of the graviton and we can see that also this contribution has its maximum amplitude in the squeezed limit.

Hence, both contributions have their maximum amplitude in the squeezed limit. If we focus our attention in cases where the mass of the graviton is small during inflation,  the first contribution is dominant and, as we have seen, it can be enhanced with respect to the similar contribution coming from standard single-field inflationary models.  Our starting theory does not show any parity violation feature at the level of the action, so we expect that $F_{{\cal T}_{1},{\cal T}_{2}}^{---}=F_{{\cal T}_{1},{\cal T}_{2}}^{+++}$. Using the properties of the polarization tensors we can also show, from \eqref{at1} and \eqref{at2}, that $F_{{\cal T}_{1},{\cal T}_{2}}^{++-}(k_1,k_2,k_3)=F_{{\cal T}_{1},{\cal T}_{2}}^{+++}(k_1,k_2,-k_3)$, and again, since we do not have parity violation, we expect that $F_{{\cal T}_{1},{\cal T}_{2}}^{--+}=F_{{\cal T}_{1},{\cal T}_{2}}^{++-}$.
It would be interesting to find a mechanism to violate parity at the level of the action and study its features, as it happens in models that involve pseudoscalar fields~\cite{Shiraishi:2013kxa,Namba:2015gja}. 

We conclude this Section by noticing that
our findings so far are relatively straightforward to investigate in our set-up of supersolid inflation with non-minimal coupling with 
 curvature. It would be interesting to examine whether also in the original solid inflation  scenario \cite{Endlich:2012pz} there  exist 
 regimes 
 where tensor non-Gaussianity can be parametrically large, and enhanced  in  squeezed configuration.

\section{Dynamics of scalar fluctuations}
\label{sec:scalar-fluc}

After analysing the dynamics of tensor modes, we now pass to discuss some aspects of scalar fluctuations in our systems.
 Features of scalar fluctuations in models of solid and supersolid inflation have already been discussed in some length in the literature  --
 see for example~\cite{Endlich:2012pz, Endlich:2013jia}. Scalar fluctuations are characterised by  a direction dependent squeezed bispectrum, an enhanced
 scalar-tensor-tensor three-point function~\cite{Dimastrogiovanni:2014ina}, and a slow suppression of any background anisotropies during inflation~\cite{Bartolo:2013msa,Bartolo:2014xfa}. We now point out another property that distinguishes 
 our system, and that we find interesting: at leading order in an  expansion of  $\epsilon_\phi$ and $\epsilon_H$ (the parameters breaking 
  time-reparameterization and de Sitter
 symmetry) 
    the dynamics of scalar curvature fluctuation depends (mainly) on $\lambda_0$,  which is the quantity that characterizes the breaking of space-reparameterizations.
  This fact has interesting  consequences for cosmological observables, like
 the tensor-to-scalar ratio $r=\mathcal{P}_{h}/\mathcal{P}_{{\cal R}}$.
 
 \smallskip
 
 We adopt a convenient gauge choice for investigating the scalar fluctuations of the fields involved, which
 we call {\it partially unitary gauge}: we set to zero the fluctuation of the scalar $\phi$, while we allow for a non-vanishing
  scalar component  $w$ for the fluctuations of the  fields $\sigma^I$ responsible for spontaneously breaking space-reparameterizations.
At linear order, our  expansion of the fields and the spacetime metric
 around a background configuration  reads
\bea
\phi&=&\phi(t)\,, \label{pert1a}
\\
\sigma^I&=&\lambda_0\,x^I+ \frac{1}{aH}\,\frac{\lambda_0}{\sqrt{-\nabla^2}}\,\partial^I w\,,
\\
d s^2&=&-d t^2 \,\left( 1+2\,N\right)+2\,a^2\,\partial_i\,B\,d t\, d x^i+a^2\,\left[\delta_{ij}\,\left(1+2 A \right) 
\right]\, d x^i d x^j\,. \label{pert1c}
\eea
  In our gauge, the scalar fluctuations  include the modes  $N$, $B$ -- which are not dynamical, and
correspond to ADM constraints~\cite{Arnowitt:1962hi} -- the mode $A{}$, 
and
the mode $w$.  All scalar
fluctuations are dimensionless. We are especially interested to investigate the dynamics of
$A$, and how it gets affected by the presence of $w$.  Notice that in this work for simplicity we do not consider the dynamics of vector modes.

The procedure we follow to determine the action for scalar fluctuations is standard (see e.g.~\cite{Maldacena:2002vr}). We substitute our Ansatz
for the scalar fields and the metric in our initial action~\eqref{eq:tot-ac1}, and we expand up to second order in scalar fluctuations. The first order action determines the exact
 solution for the background level; using the latter, we derive  the second order action for the four scalar fluctuations around the background configuration of interest.  We discuss quadratic perturbations around the power law configuration  described in Section~\ref{subsec:powerlaw}, which contains, as special case, the expansion around the de Sitter solution studied in Section~\ref{subsec:desitt}.
  
In general, one gets an intricate quadratic action mixing $A$ and $w$. We show here that there is an interesting corner in the parameter space, characterised by $\epsilon_H\ll1$ and $\epsilon_\phi\ll1$, where the dynamics of scalar fluctuations is relatively easy to investigate.
We focus on this regime in this Section, although there might be other ranges of parameter choices that lead to interesting scenarios for the scalar sector. 
 For investigating the system it is convenient to pass to Fourier space. 
 In order to kinetically demix $A$ from $\omega$, it is useful to  work with the quantity
$\tilde \omega$, which connects $\omega$ and $A$ through the relation
 \be
\omega\,=\,\tilde \omega-\frac{12\,q_2^2\,\lambda_0^4}{k\,\left(1+q_2 \lambda_0^2\right)^2}\,A\,.
\ee
We first obtain the solution for the constraint equations 
 at leading order in $\epsilon_H$, $\epsilon_\phi$ that reads
\bea\label{constraints}
N\, &=&\frac{(1-q_2 \lambda_0^2)}{(1+q_2 \lambda_0^2)}\,\frac{\dot A}{H}
+\left(\epsilon_H-\frac{\epsilon_\phi}{\left(1+q_2 \lambda_0^2\right)} \right)\,\left(
\frac{\left(1-q_2 \lambda_0^2\right) A}{
\left(1+q_2 \lambda_0^2\right)}+\frac{a^2\,H\,\dot{\tilde \omega}}{k}
\right)
\,
+\,{\cal O}\left(\epsilon_\phi^2, \epsilon_H^2\right)\,,
 \\
 B&=&
-\frac{(1-q_2^2 \lambda_0^4)}{\left(1+q_2 \lambda_0^2\right)^2}\,\frac{A}{a^2\,H}-
\frac{12 q_2^2 \,\lambda_0^4\,\dot A}{k^2 \left(1+q_2 \lambda_0^2\right)} +{\cal O}\left(\epsilon_\phi, \epsilon_H\right)\,.
 \eea
 It is also not difficult to extend the previous solutions of the constraint equations
  at higher orders in the $\epsilon$ parameters, but for our purpouse the 
  results of the above expressions are sufficient.

Substituting the constraint conditions, we find that the second order action for the scalars with momentum  $k$ results
 \bea \label{squad1}
S_{k}^{(S)}&=& \,\frac{ \bar M_{Pl}^2}{2}\,\int \,d t \, {a^3}\,\left[Q^2_A\,\dot A^2-\frac{c_A^2\,k^2}{a^2}  A^2  
-m_A^2  \,A^2\right]\,-2 \,\Sigma\,H^2\,\bar M_{Pl}^2\, \int \,d t \, a^3\,\frac{k}{a\,H}\,
\,A  \tilde \omega
\nonumber
\\
&& +\frac{\bar M_{Pl}^2}{2}\,\int \,d t \, {a^3}
\left[Q^2_{\tilde{\omega}}\,\dot{\tilde \omega}^2-\frac{c_{\tilde{\omega}}^2\,k^2}{a^2}   \tilde \omega^2
- m_{\tilde{\omega}}^2 \tilde \omega^2\right]
\,+\,{\cal O}(\epsilon^2)\,,
\eea
where
\bea
\bar{M}_{Pl}^2&=&\left(1-q_2 \lambda_0^2\right)\,M_{Pl}^2\,,
\\
{ Q}^2_A &=&
\frac{24\,   q_2^2 \lambda_0^4 }{\left(1+q_2 \lambda_0^2 \right)^2    }
+2 \epsilon_\phi\,\frac{\left(1-q_2 \lambda_0^2\right)}{\left(1+q_2 \lambda_0^2\right)^2}\,,
\\ \label{czeq}
c_A^2 &=&
\frac{8 \,q_2^2 \,\lambda_0^4}{(1-q_2^2 \lambda_0^4)}
+2 \epsilon_\phi\,\frac{\left(1-q_2 \lambda_0^2\right)}{\left(1+q_2 \lambda_0^2\right)^2}\,,\\
m_A^2&=&
\frac{12\,H^2}{\left(1+q_2 \lambda_0^2\right)^4\,\left(1-q_2 \lambda_0^2\right)}
\Big[ 
\epsilon_H\,\left(1+q_2 \lambda_0^2-8 q_2^2 \lambda_0^4+47 q_2^4 \lambda_0^8+87 q_2^5 \lambda_0^{10}\right)+
\nonumber\\
&&\hskip3.9cm
- \epsilon_\phi \left(1
+q_2 \lambda_0^2-5 q_2^2 \lambda_0^4+3 q_2^3 \lambda_0^6+32 q_2^4 \lambda_0^8
 \right)
\Big]\,,
\\
Q^2_{\tilde{\omega}}&=&\frac{2}{ \left(1-q_2 \lambda_0^2\right)} \left[
 \epsilon_H \left(1+q_2 \lambda_0^2\right)-\epsilon_\phi
\right]\,,
\\
c_{\tilde{\omega}}^2&=&
\frac{2}{  \left(1-q_2 \lambda_0^2\right)} \left[
 \epsilon_H \left(1+3 q_2 \lambda_0^2\right)-\epsilon_\phi
\right]\,,
\\ \nonumber
\\
\label{fmom}
m^2_{\tilde{\omega}}&=& -4\,Q^2_\omega\,H^2\,,
\\
\Sigma&=&\frac{4}{\left(1+q_2 \lambda_0^2\right)^2\, \left(1-q_2 \lambda_0^2\right)}
\,\left[
 \epsilon_H\left( 
1-q_2 \lambda_0^2-11 q_2^2 \lambda_0^4-21 q_2^3 \lambda_0^6
\right)-\epsilon_\phi \left( 
1-q_2 \lambda_0^2-8 q_2^2 \lambda_0^4
\right)
\right]\,.\nonumber
\eea

 Hence we have obtained the quadratic action for scalar fluctuations at leading order in an expansion of small parameters $\epsilon_\phi$, $\epsilon_H$ (but keeping
 $\lambda_0$ arbitrary). 
 
 \smallskip
 These results deserve various
 comments:
 \begin{itemize}
\item If we focus at zeroth order in an expansion in $\epsilon_H$, $\epsilon_\phi$, we  have only one  scalar propagating
 fluctuation,   the  mode $A$. The 
action in this case simplifies to 
\bea
S_{k}^A&=& \frac{12\,   q_2^2 \lambda_0^4 \,\bar M_{Pl}^2}{\left(1+q_2 \lambda_0^2 \right)^2    }\,\int \,d t\, {a^3}\,\left[\dot A^2
-\frac{(1+q_2 \lambda_0^2)}{3(1-q_2 \lambda_0^2)}\,
\frac{k^2}{a^2}  A^2  
\right]\,
\,+\,{\cal O}(\epsilon_\phi,\,\epsilon_H)\,.
\label{eq:scalact}
\eea
  Hence, the dynamics of $A$ is  controlled by the parameter $\lambda_0$, characterizing  the breaking of space-reparameterization. The sound speed
  $c_A$ for $A$ lies in the interval $1/3\le c_A \le 2/3$  (since $q_2 \lambda_0^2$  is limited within the interval of Eq.~\eqref{condqq2}).
 Scalar fluctuations acquire an adiabatic, almost scale-invariant spectrum, as in single-field slow-roll inflation.
  In this limit, the graviton mass goes to zero and the large-scale anisotropies die out exponentially fast, since tensor modes behave adiabatically. 
  \item  When we include first order corrections in the small parameters
   $\epsilon_H$, $\epsilon_\phi$, the system starts propagating the second scalar mode $\tilde \omega$ besides $A$. In order to avoid 
    ghost-like instabilities associated with $\tilde \omega$, we impose
   \be
   Q_\omega^2\,\ge\, 0 \hskip1cm \Rightarrow \hskip1cm \epsilon_H\, >\,\frac{\epsilon_\phi}{1+q_2\,\lambda_0^2} \hskip1cm \Rightarrow \hskip1cm m_h^2\,>\,
   \frac{8\,\epsilon_\phi\,q_2 \lambda_0^2}{1-q_2^2 \lambda_0^4} \,. 
   \ee
   Hence   we need a positive mass squared for the tensor modes.  The mode  $\tilde \omega$ is a tachyon with negative mass squared
   proportional to the Hubble parameter squared, see Eq.~\eqref{fmom}. On the other hand, such tachyonic instability is not necessarily an insurmountable problem, since
   in a large-scale limit $k\ll a H$ (where the tachyonic instability becomes important)  the coupling of $\tilde \omega$ with curvature perturbation is suppressed. 
   At smaller scales, the tachyon instability is less important. It would be interesting to study in full detail possible consequences of this two fields
   scalar action and possible phenomenological applications, for example along 
    the lines of the recent work~\cite{Achucarro:2016fby}, also including the effect of interactions 
     controlled by the action at third order in perturbations.   
   \item We can also check what happens `turning off' the  parameter controlling the breaking of space-reparameterization. When $\lambda_0=0$, at leading order in slow-roll $\epsilon_\phi \,=\, \epsilon_H $, all the terms in the action containing $\tilde \omega$
 vanish, and we are left with the standard quadratic action for $A$
 \bea \label{squad1a}
S^{A}_k&=&  M_{Pl}^2 \,\epsilon_{\phi}\,\int \,d t\, {a^3}\left(
 \dot A^2-\frac{k^2}{a^2}  A^2  
\right)\,,
\eea
  that reproduces well known results (see e.g.~\cite{Maldacena:2002vr}).  
 
 \item Another potentially interesting case can be obtained selecting $\epsilon_\phi\,=\,\epsilon_H (1+q_2 \lambda_0^2)$.
 This leads to $Q_\omega=0$,  implying that the mode $\tilde \omega$ does not propagate at leading order in perturbations (while it can acquire dynamics at higher orders in  perturbative expansion).  We will consider this special case
 in the next subsection.
 
 \item We do not study the scalar action expanded at third order in fluctuations since we expect that the study of scalar non-Gaussianity leads to results qualitatively similar to the ones already investigated in~\cite{Bartolo:2015qvr}, namely 
  a direction-dependent squeezed limit for the curvature bispectrum. It would be interesting, on the other hand, to study the amplitude of the the three-point function $\langle h\, A\,A\,\rangle$, given its potentially interesting observational consequences~\cite{Bordin:2016ruc,Jeong:2012df, Domenech:2017kno}.
 \end{itemize}

\subsection{Consequences for the tensor-to-scalar ratio}
\label{subsec:tsr}

Our findings for the dynamics of quadratic fluctuations in the scalar sector have potentially interesting ramifications for 
what respects  the  tensor-to-scalar ratio $r$. We start to explore some aspects of this interesting topic in this  
subsection. 
For definiteness, we focus on a specific case of the results we have
 found above, at 
 leading order in a  $\epsilon_\phi$, $\epsilon_H$ expansion.  Our aim it is not to study the 
 dynamics of scalar perturbations in full generality, but instead to exhibit an explicit, simple example where the value of $r$ does
 not depend on $\epsilon_H$, the parameter controlling the breaking of time-reparameterization, but on the parameter $\lambda_0$ controlling the breaking of space-translations.
 
 In general,
 the comoving curvature perturbation ${\cal R}$
  is defined in terms of the gauge invariant combination
  \be \label{defoR}
{\cal R}\,=\,A+\frac{\dot{A}-H\, N}{\epsilon_H\,H}\,.
\ee
In the partially unitary gauge we are adopting, see Eqs.~\eqref{pert1a}-\eqref{pert1c}, we solved the constraint
equations and determined an expression for the non-dynamical variable $N$ in Eq. \eqref{constraints}. For
simplicity, in this Section we make the following choice to relate $\epsilon_\phi$ to $\epsilon_H$
\be \label{co-ep-1-A}
\epsilon_\phi\,=\,\epsilon_H \left(1+q_2 \lambda_0^2\right)\,.
\ee
We also assume that the combination $q_2 \lambda_0^2$ is small, although well
 larger than $\epsilon_H$; in other words, we assume the hierarchy 
 \be \label{hieq2}
 \epsilon_H\, \ll \,q_2 \lambda_0^2\,\ll\,1\,.
 \ee
{\it A posteriori}, we will verify that this condition is the most interesting one for phenomenological applications.
 Working on this corner of parameter space  considerably simplifies the expression \eqref{constraints} for $N$, at leading order in the $\epsilon_i$ parameters.
Substituting the resulting expression for $N$ in the definition \eqref{defoR}, we obtain
\be \label{cupe3}
{\cal R}\,=\,A+\frac{2\,q_2\,\lambda_0^2}{\epsilon_H\,H\,\left(1+q_2 \lambda_0^2\right)}\,\dot{A}\,.
\ee
Then, the comoving curvature perturbation depends on $A$ and on its time derivative. In the previous Section, we learned that, at quadratic order, scalar fluctuations are     governed
 by the  action \eqref{squad1}.  Making the choice \eqref{co-ep-1-A}, the perturbation $\tilde \omega$
 does not propagate, and then can be integrated out: the resulting action describes a single  fluctuation $A$ with non-vanishing mass, propagating in a
 quasi-de Sitter space. The equations of motion for $A$  can then be solved exactly in terms of Hankel functions and, at large
 scales $k/a H\ll1$, we find that the solution satisfies the relation
\be \label{timeA}
\dot{A}\,=\,-\left( 1+\frac{28}{3}\,q_2\lambda_0^2
 \right)\,\epsilon_H\,H\,A\,,
\ee
which is valid at leading order in $\epsilon_H$, and up to first order in an expansion in $q_2\lambda_0^2$.  Substituting
this result~\eqref{timeA} in Eq.~\eqref{cupe3}, we find in the same regime the following proportionality relation between ${\cal R}$
and $A$ at large scales
\be
{\cal R}\,=\,\Big(1-2 \,q_2\,\lambda_0^2
\Big) \,A\,.
\ee
We now have all the ingredients
to compute the power spectrum for the comoving curvature perturbation ${\cal R}$, following the same steps
 as done for the tensor spectrum. At large scales we
 find 
 \be
 {\cal P}_{\cal R}\,=\,\frac{\sqrt{3}\,H^2}{32\, \pi^2\,\bar M_{Pl}^2}\,
\frac{1}{   q_2^2 \lambda_0^4}\,,
 \ee
plus corrections that are small in the limit of small $\epsilon_H$ and $q_2 \lambda_0^2$. Hence, we learn that the leading contribution to the amplitude of the large scale power spectrum for ${\cal R}$ does not depend on $\epsilon_H$, but on the combination $q_2 \lambda_0^2$ controlling the breaking of space-reparameterizations.

On the other hand, we recall that the amplitude for the tensor power spectrum is~\eqref{eq:tensps}
\be
{\cal P}_h\,=\,\frac{2H^2}{\pi^2\, \bar M_{Pl}^2 }\frac{\left(1- q_2\,\lambda_0^2\right)^{3/2}}{
\left(
1- 3 q_2\,\lambda_0^2\right)^{3/2}}\,\simeq\,\frac{2H^2}{\pi^2\, \bar M_{Pl}^2 } \,,
\ee
where  the second equality is valid at   zeroth order in an expansion in $q_2\,\lambda_0^2$. 

Collecting these results, we find that, at leading (zeroth) order in an expansion in $\epsilon_\phi$, $\epsilon_H$, 
and at leading order in $q_2\,\lambda_0^2$,	the tensor-to-scalar ratio reads
\be
r\,\equiv\,\frac{{\cal P}_h}{{\cal P}_{\cal R}}\,=\,
\frac{64}{\sqrt{3}}
\,
{\,q_2^2\,\lambda_0^4}
\,.
\ee
Interestingly, for our choice of hierarchy \eqref{hieq2}, at leading order $r$ depends only on the parameter $\lambda_0$ which spontaneously  breaks space-reparameterization,
and not at all on $\epsilon_H$. 

Assuming an upper bound on the tensor-to-scalar ratio $r\le0.1$  consistent with the latest CMB constraints~\cite{Array:2015xqh} allows to 
extract a limit on the (combination of) parameters $q_2 \lambda_0^2$
\be\label{boq2}
r\le0.1 \hskip0.5cm \Rightarrow \hskip0.5cm q_2 \lambda_0^2\lesssim0.05\,,
\ee
which motivates our choice \eqref{hieq2} of relatively small combination for $q_2 \lambda_0^2$.
As far as we are aware, this is the first example of inflationary scenario where the tensor-to-scalar ratio
is not proportional to $\epsilon_H$. 

\smallskip
It is important to emphasize
that this bound on  $q_2 \lambda_0^2$ only applies in the particular region of parameter space we examined: 
there might be other interesting regimes to investigate scalar fluctuations, without imposing a hierarchy as \eqref{hieq2}, where
the tensor-to-scalar ratio shows a different behaviour.
However, considering this  case for the moment, 
it is interesting to study the connection among our result and the issue of the Lyth bound. For single-field set-ups where only time-reparameterization is broken, the Lyth bound~\cite{Liddle:1992wi,Boubekeur:2012xn,Garcia-Bellido:2014wfa}
 relates   the tensor-to-scalar
 ratio $r$ with the field excursion $\Delta \phi$ of the inflaton field during inflation. A conservative value of $r$ of order $10^{-2}$ requires
 that the inflaton field excursion is larger than the Planck scale. Such large-field excursions are dangerous,
  since it is expected that Planck-scale quantum gravity effects can spoil the flatness of the potential (and the
  approximate global shift symmetry $\phi\to\phi+const.$) required to sustain a sufficiently large period of inflation. See e.g.~the recent
   \cite{Parameswaran:2016fqr} for a discussion in the context of string theory. 
   In our set-up, we find that there is no relation between $r$ and the excursion of the field $\phi$: large
  values of $r$ are compatible with sub-Planckian values of $\phi$, hence avoiding the problem. On the other hand, we
  do have field excursions on the space-like directions $x^I$, associated with $\sigma^I$. 
  It is not clear to us whether  
  spatial field excursions should 
 be limited in extensions by  some versions or generalisations of the Lyth bound, if one again wishes to avoid
  symmetry breaking  induced by quantum gravity effects.   
  We plan to investigate this topic in a separate paper, also applying the findings of the recent work~\cite{Klaewer:2016kiy}. 
  
\section{Conclusions}
\label{sec:conc}

In this paper we examined the dynamics of cosmological fluctuations in  concrete scenarios of supersolid inflation, a framework which spontaneously breaks both space and time reparameterization invariance through the vacuum expectation values of scalar fields driving inflation. We have included a non-minimal coupling of 
 scalar fields with gravity. 
Our main motivation was to show that this framework 
 can provide qualitatively new features for the dynamics of cosmological fluctuations,  which can not be reproduced in frameworks that do not break 
space-reparameterization, and that can lead to new ways to test the pattern of symmetry breaking during inflation.

We focussed in particular  on the tensor sector,  including  for the first time in this context  an analysis  of tensor non-Gaussianity. In these scenarios,
    tensor modes can have a non vanishing
  mass and a non trivial sound speed. This confirms  that primordial gravitational waves can have a blue spectrum, which make them easier to detect at smaller scales~\cite{Bartolo:2016ami}. Tensor non-Gaussianity have also distinctive
  features specifically associated with the  
  pattern of symmetry breaking that we have considered. The third order action for tensor modes has two contributions: one with the same structure as the usual one derived from General Relativity (but with a different overall coefficient), the other is  new  and specific of our simmetry breaking set-up.  We found that the tensor bispectrum is peaked in the squeezed limit, with an amplitude  that can be parametrically larger than in standard single-field scenarios of inflation.  It would be interesting to investigate whether a large amplitude
  for the squeezed limit  of tensor bispectrum can facilitate the detection of tensor non-Gaussianity, for example inducing large scale anisotropies in the tensor power spectrum, analogously to what happens in the scalar case.
  
  We then analysed the dynamics of scalar perturbations. In general, in this kind of scenarios, two scalar modes propagate and we found that the
  analysis simplifies considerably at leading  order in an expansion in the small parameters
  breaking time-reparameterization and the de Sitter symmetry -- while keeping a larger size for the parameter controlling
the  breaking of space-reparameterization. At leading order in such expansion, only one scalar mode propagate, which at large scales can be  identified
with the comoving curvature perturbation ${\cal R}$.
 At next to leading order, a second mode becomes dynamical, with a tachyonic instability 
whose effects can be suppressed by the small expansion parameters. 

The fact that, at leading order in our expansion, 
   the amplitude of curvature fluctuations is dictated by the parameter controlling
the  breaking of space-reparameterization is an interesting feature of our set up.  
 As a consequence, in this regime the tensor-to-scalar ratio $r$ is independent of the parameter
$\epsilon_H$ which controls the breaking of time reparameterization  during inflation, as usually happens.  Instead, in our case, for the first
time we determine scenarios where $r$ depends 
 on quantities controlling the breaking of space-reparameterization. 
 It would be interesting to  investigate
 in more details the consequences of these findings for the effective
field theory of inflation, and for issues related to trans-Planckian field excursions and the Lyth bound. 

\subsection*{Acknowledgments}
We thank Ivonne Zavala for careful reading of the manuscript. 
The computations performed in this paper have been partially done with the \textit{xAct}
package for Mathematica~\cite{xact,Brizuela:2008ra,Pitrou:2013hga}.

\begin{appendix}
\section{Appendix A}\label{app-A}
In this appendix we collect some results useful for expanding the action at second and third order in tensor fluctuations
 \bea
\sqrt{-g}&=&a^3\,,
\\
\sqrt{-g} \, g_{ij} \delta^{ij}&=&3 \,a+\frac{a}{2}\,h_{im}\,h_{mj}-\frac{a}{6} h_{i m} h_{m l} h_{lj}\,,
\\
\sqrt{-g} \, R&=&a^3\Big[ 12 H^2+6 \dot{H}+\frac14\left( \dot{h}_{ij}^2 -\frac{1}{a^2} \left(\partial_m h_{ij}\right)^2\right)+
\nonumber\\&&
+\frac{1}{2\,a^2} h_{ij} h_{mn} \left( 
\partial_j \partial_n h_{im}-\frac12 \partial_i \partial_j h_{mn}
\right)\Big]\,,
\\
\sqrt{-g}\,G_{ij}\,\delta^{ij}&=&-3 a\left(3 H^2+2 \dot{H}\right)+
\nonumber\\
&&-\frac{a}{8}
\left[ 3 \dot{h}_{ij} \left( \dot{h}_{ij}+4 H h_{ij}\right) +4 h_{ij} \left( \ddot{h}_{ij}+h_{ij} \left(3 H^2+2 \dot{H}\right) \right)
+\frac{3}{a^2} \left( \partial_l h_{ij}\right)^2\right]
\nonumber\\
&&
+{a} \left(\frac14 \, h_{ij} h_{jm} \ddot{h}_{mi}+\frac{3}{4}  H h_{ij} \dot{h}_{jm} h_{mi}+\frac{\dot{H}}{3} h_{ij} h_{jm} h_{mi}
+\frac{H^2}{2} h_{ij} h_{jm} h_{mi} \right)
\nonumber\\
&&
+\frac{5}{4 a} h_{ij} h_{mn}\left( \partial_j \partial_n h_{im}-\frac12 \partial_i \partial_j h_{mn}
\right)-\frac{1}{4 a} h_{ij} h_{im} \nabla^2 h_{mj}\,.
\eea
Latin indexes have been contracted with the 3d Kronecker symbol $\delta_{ij}$.

\end{appendix}

\bibliographystyle{utcaps}
\bibliography{TensBib.bib}

\providecommand{\href}[2]{#2}\begingroup\raggedright\begin{thebibliography}{10}

\bibitem{Cheung:2007st}
C.~Cheung, P.~Creminelli, A.~L. Fitzpatrick, J.~Kaplan, and L.~Senatore, ``{The
  Effective Field Theory of Inflation},''
  \href{http://dx.doi.org/10.1088/1126-6708/2008/03/014}{{\em JHEP} {\bfseries
  0803} (2008) 014},
\href{http://arxiv.org/abs/0709.0293}{{\ttfamily arXiv:0709.0293 [hep-th]}}.

\bibitem{Weinberg:2008hq}
S.~Weinberg, ``{Effective Field Theory for Inflation},''
  \href{http://dx.doi.org/10.1103/PhysRevD.77.123541}{{\em Phys. Rev.}
  {\bfseries D77} (2008) 123541},
\href{http://arxiv.org/abs/0804.4291}{{\ttfamily arXiv:0804.4291 [hep-th]}}.

\bibitem{Piazza:2013coa}
F.~Piazza and F.~Vernizzi, ``{Effective Field Theory of Cosmological
  Perturbations},''
  \href{http://dx.doi.org/10.1088/0264-9381/30/21/214007}{{\em Class. Quant.
  Grav.} {\bfseries 30} (2013) 214007},
\href{http://arxiv.org/abs/1307.4350}{{\ttfamily arXiv:1307.4350 [hep-th]}}.

\bibitem{Golovnev:2008cf}
A.~Golovnev, V.~Mukhanov, and V.~Vanchurin, ``{Vector Inflation},''
  \href{http://dx.doi.org/10.1088/1475-7516/2008/06/009}{{\em JCAP} {\bfseries
  0806} (2008) 009},
\href{http://arxiv.org/abs/0802.2068}{{\ttfamily arXiv:0802.2068 [astro-ph]}}.

\bibitem{Endlich:2012pz}
S.~Endlich, A.~Nicolis, and J.~Wang, ``{Solid Inflation},''
  \href{http://dx.doi.org/10.1088/1475-7516/2013/10/011}{{\em JCAP} {\bfseries
  1310} (2013) 011},
\href{http://arxiv.org/abs/1210.0569}{{\ttfamily arXiv:1210.0569 [hep-th]}}.

\bibitem{Gruzinov:2004ty}
A.~Gruzinov, ``{Elastic inflation},''
  \href{http://dx.doi.org/10.1103/PhysRevD.70.063518}{{\em Phys. Rev.}
  {\bfseries D70} (2004) 063518},
\href{http://arxiv.org/abs/astro-ph/0404548}{{\ttfamily arXiv:astro-ph/0404548
  [astro-ph]}}.

\bibitem{Himmetoglu:2008zp}
B.~Himmetoglu, C.~R. Contaldi, and M.~Peloso, ``{Instability of anisotropic
  cosmological solutions supported by vector fields},''
  \href{http://dx.doi.org/10.1103/PhysRevLett.102.111301}{{\em Phys. Rev.
  Lett.} {\bfseries 102} (2009) 111301},
\href{http://arxiv.org/abs/0809.2779}{{\ttfamily arXiv:0809.2779 [astro-ph]}}.

\bibitem{Maleknejad:2011jw}
A.~Maleknejad and M.~M. Sheikh-Jabbari, ``{Gauge-flation: Inflation From
  Non-Abelian Gauge Fields},''
  \href{http://dx.doi.org/10.1016/j.physletb.2013.05.001}{{\em Phys. Lett.}
  {\bfseries B723} (2013) 224--228},
\href{http://arxiv.org/abs/1102.1513}{{\ttfamily arXiv:1102.1513 [hep-ph]}}.

\bibitem{Maleknejad:2011sq}
A.~Maleknejad and M.~M. Sheikh-Jabbari, ``{Non-Abelian Gauge Field
  Inflation},'' \href{http://dx.doi.org/10.1103/PhysRevD.84.043515}{{\em Phys.
  Rev.} {\bfseries D84} (2011) 043515},
\href{http://arxiv.org/abs/1102.1932}{{\ttfamily arXiv:1102.1932 [hep-ph]}}.

\bibitem{Adshead:2012kp}
P.~Adshead and M.~Wyman, ``{Chromo-Natural Inflation: Natural inflation on a
  steep potential with classical non-Abelian gauge fields},''
  \href{http://dx.doi.org/10.1103/PhysRevLett.108.261302}{{\em Phys. Rev.
  Lett.} {\bfseries 108} (2012) 261302},
\href{http://arxiv.org/abs/1202.2366}{{\ttfamily arXiv:1202.2366 [hep-th]}}.

\bibitem{Maleknejad:2012fw}
A.~Maleknejad, M.~M. Sheikh-Jabbari, and J.~Soda, ``{Gauge Fields and
  Inflation},'' \href{http://dx.doi.org/10.1016/j.physrep.2013.03.003}{{\em
  Phys. Rept.} {\bfseries 528} (2013) 161--261},
\href{http://arxiv.org/abs/1212.2921}{{\ttfamily arXiv:1212.2921 [hep-th]}}.

\bibitem{Bartolo:2012sd}
N.~Bartolo, S.~Matarrese, M.~Peloso, and A.~Ricciardone, ``{Anisotropic power
  spectrum and bispectrum in the $f(\phi)F^2$ mechanism},''
  \href{http://dx.doi.org/10.1103/PhysRevD.87.023504}{{\em Phys. Rev.}
  {\bfseries D87} no.~2, (2013) 023504},
\href{http://arxiv.org/abs/1210.3257}{{\ttfamily arXiv:1210.3257
  [astro-ph.CO]}}.

\bibitem{Son:2005ak}
D.~T. Son, ``{Effective Lagrangian and topological interactions in
  supersolids},'' \href{http://dx.doi.org/10.1103/PhysRevLett.94.175301}{{\em
  Phys. Rev. Lett.} {\bfseries 94} (2005) 175301},
\href{http://arxiv.org/abs/cond-mat/0501658}{{\ttfamily arXiv:cond-mat/0501658
  [cond-mat]}}.

\bibitem{Bartolo:2015qvr}
N.~Bartolo, D.~Cannone, A.~Ricciardone, and G.~Tasinato, ``{Distinctive
  signatures of space-time diffeomorphism breaking in EFT of inflation},''
  \href{http://dx.doi.org/10.1088/1475-7516/2016/03/044}{{\em JCAP} {\bfseries
  1603} no.~03, (2016) 044},
\href{http://arxiv.org/abs/1511.07414}{{\ttfamily arXiv:1511.07414
  [astro-ph.CO]}}.

\bibitem{AmaroSeoane:2012km}
P.~Amaro-Seoane {\em et~al.}, ``{eLISA/NGO: Astrophysics and cosmology in the
  gravitational-wave millihertz regime},'' {\em GW Notes} {\bfseries 6} (2013)
  4--110,
\href{http://arxiv.org/abs/1201.3621}{{\ttfamily arXiv:1201.3621
  [astro-ph.CO]}}.

\bibitem{Koh:2013msa}
S.~Koh, S.~Kouwn, O.-K. Kwon, and P.~Oh, ``{Cosmological Perturbations of a
  Quartet of Scalar Fields with a Spatially Constant Gradient},''
  \href{http://dx.doi.org/10.1103/PhysRevD.88.043523}{{\em Phys. Rev.}
  {\bfseries D88} (2013) 043523},
\href{http://arxiv.org/abs/1304.7924}{{\ttfamily arXiv:1304.7924 [gr-qc]}}.

\bibitem{Horndeski:1974wa}
G.~W. Horndeski, ``{Second-order scalar-tensor field equations in a
  four-dimensional space},''
\href{http://dx.doi.org/10.1007/BF01807638}{{\em Int. J. Theor. Phys.}
  {\bfseries 10} (1974) 363--384}.

\bibitem{Copeland:2012qf}
E.~J. Copeland, A.~Padilla, and P.~M. Saffin, ``{The cosmology of the
  Fab-Four},'' \href{http://dx.doi.org/10.1088/1475-7516/2012/12/026}{{\em
  JCAP} {\bfseries 1212} (2012) 026},
\href{http://arxiv.org/abs/1208.3373}{{\ttfamily arXiv:1208.3373 [hep-th]}}.

\bibitem{Kobayashi:2010cm}
T.~Kobayashi, M.~Yamaguchi, and J.~Yokoyama, ``{G-inflation: Inflation driven
  by the Galileon field},''
  \href{http://dx.doi.org/10.1103/PhysRevLett.105.231302}{{\em Phys. Rev.
  Lett.} {\bfseries 105} (2010) 231302},
\href{http://arxiv.org/abs/1008.0603}{{\ttfamily arXiv:1008.0603 [hep-th]}}.

\bibitem{Kobayashi:2011nu}
T.~Kobayashi, M.~Yamaguchi, and J.~Yokoyama, ``{Generalized G-inflation:
  Inflation with the most general second-order field equations},''
  \href{http://dx.doi.org/10.1143/PTP.126.511}{{\em Prog. Theor. Phys.}
  {\bfseries 126} (2011) 511--529},
\href{http://arxiv.org/abs/1105.5723}{{\ttfamily arXiv:1105.5723 [hep-th]}}.

\bibitem{Watanabe:2009ct}
M.-a. Watanabe, S.~Kanno, and J.~Soda, ``{Inflationary Universe with
  Anisotropic Hair},''
  \href{http://dx.doi.org/10.1103/PhysRevLett.102.191302}{{\em Phys. Rev.
  Lett.} {\bfseries 102} (2009) 191302},
\href{http://arxiv.org/abs/0902.2833}{{\ttfamily arXiv:0902.2833 [hep-th]}}.

\bibitem{Abolhasani:2015cve}
A.~A. Abolhasani, M.~Akhshik, R.~Emami, and H.~Firouzjahi, ``{Primordial
  Statistical Anisotropies: The Effective Field Theory Approach},''
  \href{http://dx.doi.org/10.1088/1475-7516/2016/03/020}{{\em JCAP} {\bfseries
  1603} no.~03, (2016) 020},
\href{http://arxiv.org/abs/1511.03218}{{\ttfamily arXiv:1511.03218
  [astro-ph.CO]}}.

\bibitem{Lucchin:1984yf}
F.~Lucchin and S.~Matarrese, ``{Power Law Inflation},''
\href{http://dx.doi.org/10.1103/PhysRevD.32.1316}{{\em Phys. Rev.} {\bfseries
  D32} (1985) 1316}.

\bibitem{Liddle:1988tb}
A.~R. Liddle, ``{Power Law Inflation With Exponential Potentials},''
\href{http://dx.doi.org/10.1016/0370-2693(89)90776-4}{{\em Phys. Lett.}
  {\bfseries B220} (1989) 502--508}.

\bibitem{Cannone:2014uqa}
D.~Cannone, G.~Tasinato, and D.~Wands, ``{Generalised tensor fluctuations and
  inflation},''
\href{http://arxiv.org/abs/1409.6568}{{\ttfamily arXiv:1409.6568
  [astro-ph.CO]}}.

\bibitem{Cannone:2015rra}
D.~Cannone, J.-O. Gong, and G.~Tasinato, ``{Breaking discrete symmetries in the
  effective field theory of inflation},''
  \href{http://dx.doi.org/10.1088/1475-7516/2015/08/003}{{\em JCAP} {\bfseries
  1508} no.~08, (2015) 003},
\href{http://arxiv.org/abs/1505.05773}{{\ttfamily arXiv:1505.05773 [hep-th]}}.

\bibitem{Maldacena:2002vr}
J.~M. Maldacena, ``{Non-Gaussian features of primordial fluctuations in single
  field inflationary models},''
  \href{http://dx.doi.org/10.1088/1126-6708/2003/05/013}{{\em JHEP} {\bfseries
  0305} (2003) 013},
\href{http://arxiv.org/abs/astro-ph/0210603}{{\ttfamily arXiv:astro-ph/0210603
  [astro-ph]}}.

\bibitem{Higuchi:1986py}
A.~Higuchi, ``{Forbidden Mass Range for Spin-2 Field Theory in De Sitter
  Space-time},''
\href{http://dx.doi.org/10.1016/0550-3213(87)90691-2}{{\em Nucl. Phys.}
  {\bfseries B282} (1987) 397--436}.

\bibitem{Bartolo:2013msa}
N.~Bartolo, S.~Matarrese, M.~Peloso, and A.~Ricciardone, ``{Anisotropy in solid
  inflation},'' \href{http://dx.doi.org/10.1088/1475-7516/2013/08/022}{{\em
  JCAP} {\bfseries 1308} (2013) 022},
\href{http://arxiv.org/abs/1306.4160}{{\ttfamily arXiv:1306.4160
  [astro-ph.CO]}}.

\bibitem{Bartolo:2014xfa}
N.~Bartolo, M.~Peloso, A.~Ricciardone, and C.~Unal, ``{The expected anisotropy
  in solid inflation},''
  \href{http://dx.doi.org/10.1088/1475-7516/2014/11/009}{{\em JCAP} {\bfseries
  1411} no.~11, (2014) 009},
\href{http://arxiv.org/abs/1407.8053}{{\ttfamily arXiv:1407.8053
  [astro-ph.CO]}}.

\bibitem{Akhshik:2014bla}
M.~Akhshik, ``{Clustering Fossils in Solid Inflation},''
  \href{http://dx.doi.org/10.1088/1475-7516/2015/05/043}{{\em JCAP} {\bfseries
  1505} no.~05, (2015) 043},
\href{http://arxiv.org/abs/1409.3004}{{\ttfamily arXiv:1409.3004
  [astro-ph.CO]}}.

\bibitem{Akhshik:2014gja}
M.~Akhshik, R.~Emami, H.~Firouzjahi, and Y.~Wang, ``{Statistical Anisotropies
  in Gravitational Waves in Solid Inflation},''
  \href{http://dx.doi.org/10.1088/1475-7516/2014/09/012}{{\em JCAP} {\bfseries
  1409} (2014) 012},
\href{http://arxiv.org/abs/1405.4179}{{\ttfamily arXiv:1405.4179
  [astro-ph.CO]}}.

\bibitem{Lyth:2009zz}
D.~H. Lyth and A.~R. Liddle, {\em {The primordial density perturbation:
  Cosmology, inflation and the origin of structure}}.
\newblock 2009.
\newblock
\url{http://www.cambridge.org/uk/catalogue/catalogue.asp?isbn=9780521828499}.
\newblock

\bibitem{Gao:2011vs}
X.~Gao, T.~Kobayashi, M.~Yamaguchi, and J.~Yokoyama, ``{Primordial
  non-Gaussianities of gravitational waves in the most general single-field
  inflation model},''
  \href{http://dx.doi.org/10.1103/PhysRevLett.107.211301}{{\em Phys. Rev.
  Lett.} {\bfseries 107} (2011) 211301},
\href{http://arxiv.org/abs/1108.3513}{{\ttfamily arXiv:1108.3513
  [astro-ph.CO]}}.

\bibitem{Green:2009ds}
D.~Green, B.~Horn, L.~Senatore, and E.~Silverstein, ``{Trapped Inflation},''
  \href{http://dx.doi.org/10.1103/PhysRevD.80.063533}{{\em Phys. Rev.}
  {\bfseries D80} (2009) 063533},
\href{http://arxiv.org/abs/0902.1006}{{\ttfamily arXiv:0902.1006 [hep-th]}}.

\bibitem{Anber:2009ua}
M.~M. Anber and L.~Sorbo, ``{Naturally inflating on steep potentials through
  electromagnetic dissipation},''
  \href{http://dx.doi.org/10.1103/PhysRevD.81.043534}{{\em Phys. Rev.}
  {\bfseries D81} (2010) 043534},
\href{http://arxiv.org/abs/0908.4089}{{\ttfamily arXiv:0908.4089 [hep-th]}}.

\bibitem{Barnaby:2010vf}
N.~Barnaby and M.~Peloso, ``{Large Nongaussianity in Axion Inflation},''
  \href{http://dx.doi.org/10.1103/PhysRevLett.106.181301}{{\em Phys. Rev.
  Lett.} {\bfseries 106} (2011) 181301},
\href{http://arxiv.org/abs/1011.1500}{{\ttfamily arXiv:1011.1500 [hep-ph]}}.

\bibitem{Biagetti:2013kwa}
M.~Biagetti, M.~Fasiello, and A.~Riotto, ``{Enhancing Inflationary Tensor Modes
  through Spectator Fields},''
  \href{http://dx.doi.org/10.1103/PhysRevD.88.103518}{{\em Phys. Rev.}
  {\bfseries D88} (2013) 103518},
\href{http://arxiv.org/abs/1305.7241}{{\ttfamily arXiv:1305.7241
  [astro-ph.CO]}}.

\bibitem{Biagetti:2014asa}
M.~Biagetti, E.~Dimastrogiovanni, M.~Fasiello, and M.~Peloso, ``{Gravitational
  Waves and Scalar Perturbations from Spectator Fields},''
  \href{http://dx.doi.org/10.1088/1475-7516/2015/04/011}{{\em JCAP} {\bfseries
  1504} (2015) 011},
\href{http://arxiv.org/abs/1411.3029}{{\ttfamily arXiv:1411.3029
  [astro-ph.CO]}}.

\bibitem{Guzzetti:2016mkm}
C.~Guzzetti, M., N.~Bartolo, M.~Liguori, and S.~Matarrese, ``{Gravitational
  waves from inflation},''
  \href{http://dx.doi.org/10.1393/ncr/i2016-10127-1}{{\em Riv. Nuovo Cim.}
  {\bfseries 39} no.~9, (2016) 399--495},
\href{http://arxiv.org/abs/1605.01615}{{\ttfamily arXiv:1605.01615
  [astro-ph.CO]}}.

\bibitem{Bartolo:2016ami}
N.~Bartolo {\em et~al.}, ``{Science with the space-based interferometer LISA.
  IV: Probing inflation with gravitational waves},''
\href{http://arxiv.org/abs/1610.06481}{{\ttfamily arXiv:1610.06481
  [astro-ph.CO]}}.

\bibitem{Creminelli:2014wna}
P.~Creminelli, J.~Gleyzes, J.~Nore{\~n}a, and F.~Vernizzi, ``{Resilience of the
  standard predictions for primordial tensor modes},''
  \href{http://dx.doi.org/10.1103/PhysRevLett.113.231301}{{\em Phys. Rev.
  Lett.} {\bfseries 113} no.~23, (2014) 231301},
\href{http://arxiv.org/abs/1407.8439}{{\ttfamily arXiv:1407.8439
  [astro-ph.CO]}}.

\bibitem{Fumagalli:2016afy}
J.~Fumagalli, S.~Mooij, and M.~Postma, ``{Tensor power spectrum and disformal
  transformations},''
\href{http://arxiv.org/abs/1610.08460}{{\ttfamily arXiv:1610.08460 [gr-qc]}}.

\bibitem{Bekenstein:1992pj}
J.~D. Bekenstein, ``{The Relation between physical and gravitational
  geometry},'' \href{http://dx.doi.org/10.1103/PhysRevD.48.3641}{{\em Phys.
  Rev.} {\bfseries D48} (1993) 3641--3647},
\href{http://arxiv.org/abs/gr-qc/9211017}{{\ttfamily arXiv:gr-qc/9211017
  [gr-qc]}}.

\bibitem{Maldacena:2011nz}
J.~M. Maldacena and G.~L. Pimentel, ``{On graviton non-Gaussianities during
  inflation},'' \href{http://dx.doi.org/10.1007/JHEP09(2011)045}{{\em JHEP}
  {\bfseries 09} (2011) 045},
\href{http://arxiv.org/abs/1104.2846}{{\ttfamily arXiv:1104.2846 [hep-th]}}.

\bibitem{Soda:2011am}
J.~Soda, H.~Kodama, and M.~Nozawa, ``{Parity Violation in Graviton
  Non-gaussianity},'' \href{http://dx.doi.org/10.1007/JHEP08(2011)067}{{\em
  JHEP} {\bfseries 08} (2011) 067},
\href{http://arxiv.org/abs/1106.3228}{{\ttfamily arXiv:1106.3228 [hep-th]}}.

\bibitem{Cook:2013xea}
J.~L. Cook and L.~Sorbo, ``{An inflationary model with small scalar and large
  tensor nongaussianities},''
  \href{http://dx.doi.org/10.1088/1475-7516/2013/11/047}{{\em JCAP} {\bfseries
  1311} (2013) 047},
\href{http://arxiv.org/abs/1307.7077}{{\ttfamily arXiv:1307.7077
  [astro-ph.CO]}}.

\bibitem{Akita:2015mho}
Y.~Akita and T.~Kobayashi, ``{Primordial non-Gaussianities of gravitational
  waves beyond Horndeski theories},''
  \href{http://dx.doi.org/10.1103/PhysRevD.93.043519}{{\em Phys. Rev.}
  {\bfseries D93} no.~4, (2016) 043519},
\href{http://arxiv.org/abs/1512.01380}{{\ttfamily arXiv:1512.01380 [hep-th]}}.

\bibitem{Weinberg:2005vy}
S.~Weinberg, ``{Quantum contributions to cosmological correlations},''
  \href{http://dx.doi.org/10.1103/PhysRevD.72.043514}{{\em Phys. Rev.}
  {\bfseries D72} (2005) 043514},
\href{http://arxiv.org/abs/hep-th/0506236}{{\ttfamily arXiv:hep-th/0506236
  [hep-th]}}.

\bibitem{Shiraishi:2013kxa}
M.~Shiraishi, A.~Ricciardone, and S.~Saga, ``{Parity violation in the CMB
  bispectrum by a rolling pseudoscalar},''
  \href{http://dx.doi.org/10.1088/1475-7516/2013/11/051}{{\em JCAP} {\bfseries
  1311} (2013) 051},
\href{http://arxiv.org/abs/1308.6769}{{\ttfamily arXiv:1308.6769
  [astro-ph.CO]}}.

\bibitem{Namba:2015gja}
R.~Namba, M.~Peloso, M.~Shiraishi, L.~Sorbo, and C.~Unal, ``{Scale-dependent
  gravitational waves from a rolling axion},''
  \href{http://dx.doi.org/10.1088/1475-7516/2016/01/041}{{\em JCAP} {\bfseries
  1601} no.~01, (2016) 041},
\href{http://arxiv.org/abs/1509.07521}{{\ttfamily arXiv:1509.07521
  [astro-ph.CO]}}.

\bibitem{Endlich:2013jia}
S.~Endlich, B.~Horn, A.~Nicolis, and J.~Wang, ``{Squeezed limit of the solid
  inflation three-point function},''
  \href{http://dx.doi.org/10.1103/PhysRevD.90.063506}{{\em Phys. Rev.}
  {\bfseries D90} no.~6, (2014) 063506},
\href{http://arxiv.org/abs/1307.8114}{{\ttfamily arXiv:1307.8114 [hep-th]}}.

\bibitem{Dimastrogiovanni:2014ina}
E.~Dimastrogiovanni, M.~Fasiello, D.~Jeong, and M.~Kamionkowski,
  ``{Inflationary tensor fossils in large-scale structure},''
  \href{http://dx.doi.org/10.1088/1475-7516/2014/12/050}{{\em JCAP} {\bfseries
  1412} (2014) 050},
\href{http://arxiv.org/abs/1407.8204}{{\ttfamily arXiv:1407.8204
  [astro-ph.CO]}}.

\bibitem{Arnowitt:1962hi}
R.~L. Arnowitt, S.~Deser, and C.~W. Misner, ``{The Dynamics of general
  relativity},'' \href{http://dx.doi.org/10.1007/s10714-008-0661-1}{{\em Gen.
  Rel. Grav.} {\bfseries 40} (2008) 1997--2027},
\href{http://arxiv.org/abs/gr-qc/0405109}{{\ttfamily arXiv:gr-qc/0405109
  [gr-qc]}}.

\bibitem{Achucarro:2016fby}
A.~Ach{\'u}carro, V.~Atal, C.~Germani, and G.~A. Palma, ``{Cumulative effects
  in inflation with ultra-light entropy modes},''
\href{http://arxiv.org/abs/1607.08609}{{\ttfamily arXiv:1607.08609
  [astro-ph.CO]}}.

\bibitem{Bordin:2016ruc}
L.~Bordin, P.~Creminelli, M.~Mirbabayi, and J.~Norena, ``{Tensor Squeezed
  Limits and the Higuchi Bound},''
  \href{http://dx.doi.org/10.1088/1475-7516/2016/09/041}{{\em JCAP} {\bfseries
  1609} no.~09, (2016) 041},
\href{http://arxiv.org/abs/1605.08424}{{\ttfamily arXiv:1605.08424
  [astro-ph.CO]}}.

\bibitem{Jeong:2012df}
D.~Jeong and M.~Kamionkowski, ``{Clustering Fossils from the Early Universe},''
  \href{http://dx.doi.org/10.1103/PhysRevLett.108.251301}{{\em Phys. Rev.
  Lett.} {\bfseries 108} (2012) 251301},
\href{http://arxiv.org/abs/1203.0302}{{\ttfamily arXiv:1203.0302
  [astro-ph.CO]}}.

\bibitem{Domenech:2017kno}
G.~Dom{\`e}nech, T.~Hiramatsu, C.~Lin, M.~Sasaki, M.~Shiraishi, and Y.~Wang,
  ``{CMB Scale Dependent Non-Gaussianity from Massive Gravity during
  Inflation},''
\href{http://arxiv.org/abs/1701.05554}{{\ttfamily arXiv:1701.05554
  [astro-ph.CO]}}.

\bibitem{Array:2015xqh}
{\bfseries BICEP2, Keck Array} Collaboration, P.~A.~R. Ade {\em et~al.},
  ``{Improved Constraints on Cosmology and Foregrounds from BICEP2 and Keck
  Array Cosmic Microwave Background Data with Inclusion of 95 GHz Band},''
  \href{http://dx.doi.org/10.1103/PhysRevLett.116.031302}{{\em Phys. Rev.
  Lett.} {\bfseries 116} (2016) 031302},
\href{http://arxiv.org/abs/1510.09217}{{\ttfamily arXiv:1510.09217
  [astro-ph.CO]}}.

\bibitem{Liddle:1992wi}
A.~R. Liddle and D.~H. Lyth, ``{COBE, gravitational waves, inflation and
  extended inflation},''
  \href{http://dx.doi.org/10.1016/0370-2693(92)91393-N}{{\em Phys. Lett.}
  {\bfseries B291} (1992) 391--398},
\href{http://arxiv.org/abs/astro-ph/9208007}{{\ttfamily arXiv:astro-ph/9208007
  [astro-ph]}}.

\bibitem{Boubekeur:2012xn}
L.~Boubekeur, ``{Theoretical bounds on the tensor-to-scalar ratio in the cosmic
  microwave background},''
  \href{http://dx.doi.org/10.1103/PhysRevD.87.061301}{{\em Phys. Rev.}
  {\bfseries D87} no.~6, (2013) 061301},
\href{http://arxiv.org/abs/1208.0210}{{\ttfamily arXiv:1208.0210
  [astro-ph.CO]}}.

\bibitem{Garcia-Bellido:2014wfa}
J.~Garcia-Bellido, D.~Roest, M.~Scalisi, and I.~Zavala, ``{Lyth bound of
  inflation with a tilt},''
  \href{http://dx.doi.org/10.1103/PhysRevD.90.123539}{{\em Phys. Rev.}
  {\bfseries D90} no.~12, (2014) 123539},
\href{http://arxiv.org/abs/1408.6839}{{\ttfamily arXiv:1408.6839 [hep-th]}}.

\bibitem{Parameswaran:2016fqr}
S.~L. Parameswaran and I.~Zavala, ``{Prospects for Primordial Gravitational
  Waves in String Inflation},''
\href{http://arxiv.org/abs/1606.02537}{{\ttfamily arXiv:1606.02537 [hep-th]}}.

\bibitem{Klaewer:2016kiy}
D.~Klaewer and E.~Palti, ``{Super-Planckian Spatial Field Variations and
  Quantum Gravity},''
\href{http://arxiv.org/abs/1610.00010}{{\ttfamily arXiv:1610.00010 [hep-th]}}.

\bibitem{xact}
J.~M. Mart\'in-Garc\'ia, ``http://www.xact.es/.''.

\bibitem{Brizuela:2008ra}
D.~Brizuela, J.~M. Martin-Garcia, and G.~A. Mena~Marugan, ``{xPert: Computer
  algebra for metric perturbation theory},''
  \href{http://dx.doi.org/10.1007/s10714-009-0773-2}{{\em Gen. Rel. Grav.}
  {\bfseries 41} (2009) 2415--2431},
\href{http://arxiv.org/abs/0807.0824}{{\ttfamily arXiv:0807.0824 [gr-qc]}}.

\bibitem{Pitrou:2013hga}
C.~Pitrou, X.~Roy, and O.~Umeh, ``{xPand: An algorithm for perturbing
  homogeneous cosmologies},''
  \href{http://dx.doi.org/10.1088/0264-9381/30/16/165002}{{\em Class. Quant.
  Grav.} {\bfseries 30} (2013) 165002},
\href{http://arxiv.org/abs/1302.6174}{{\ttfamily arXiv:1302.6174
  [astro-ph.CO]}}.

\end{thebibliography}\endgroup


\end{document}